%% file: CompressJune14,2019.tex
\documentclass[12pt]{iopart} 

\usepackage{iopams,amsfonts,amssymb,amsthm} 

\pdfoutput=1
\usepackage{etex}
\usepackage{cite}
\usepackage{latexsym}
\usepackage{rawfonts}
\usepackage{graphicx}
\usepackage[usenames,dvipsnames,svgnames]{xcolor}
\usepackage{bbm}
\usepackage{latexsym}
\usepackage{multirow}
\usepackage{rotating}
\usepackage{lscape}
\usepackage{graphicx} 
\usepackage{subfigure}
\usepackage{xcolor}
\usepackage{fancybox}
\usepackage{ifmtarg}
\usepackage{fancyhdr}
\usepackage{wrapfig}

\input prepictex
\input pictex
\input postpictex

\usepackage{times}

\usepackage[font=footnotesize,labelfont=sf]{caption}

\include{macros}
\include{macros-pictex}

\begin{document}

\title{The free energy of compressed lattice knots}
\author{
EJ Janse van Rensburg$^1$\footnote[2]{\texttt{rensburg@yorku.ca}}
}

\address{\sf$^1$Department of Mathematics and Statistics, 
York University, Toronto, Ontario M3J~1P3, Canada\\}

\date{\today} 

\begin{abstract} \sf
A compressed knotted ring polymer in a confining cavity is
modelled by a knotted lattice polygon confined in a cube in $\IntZ^3$.  The 
GAS algorithm \cite{JvRR11} is used to sample lattice polygons of fixed knot
type in a confining cube and to estimate the free energy of confined lattice knots.
Lattice polygons of knot types the unknot, the trefoil knot, and the figure 
eight knot, are sampled and the free energies are estimated as functions of the 
concentration of monomers in the confining cube.  The data show that 
the free energy is a function of knot type at low concentrations, and 
(mean-field) Flory-Huggins theory \cite{Flory42,H42} is used 
to model the free energy as a function of monomer concentration.
The Flory interaction parameter of knotted lattice polygons in $\IntZ^3$ is also
estimated.  
\end{abstract}

\maketitle

\section{Introduction}

The effects of topology on the free energy of a polymer is a widely studied
topic \cite{MW82,deG84,TAVR01}.  The biological function of DNA and
and other biopolymers are dependent on entanglement and linking \cite{CW90}
and so living cells evolved mechanisms for manipulating knotting and 
linking \cite{KM91,RCV93,RVC97}.
  
The study of knotting in polymers dates back more than 50 years 
\cite{D62,Edwards67}.  The free energy of polymers in solvents,
and of polymer melts, and in particular the entropy of these solvents and
melts, is known to depend on the knotting, linking and entanglement
 \cite{VLKA74,MW82,deG84}.   It is also known,
both numerically and by rigorous proof in several random polygon models 
of ring polymers, that there is a high probability of knotting, at least in the 
asymptotic regime \cite{SW88,P89,J94,SW93}.  The effects of knotting and 
entanglements on the free energy of a polymer have been analysed in a variety of 
different studies and models \cite{TJvROSW94,MWC96,OW07,MO12} and 
there is now a large literature devoted to knots in models of polymers;
see for example references \cite{SSS93,SS98,MMOS06,OSV09}.

\begin{figure}
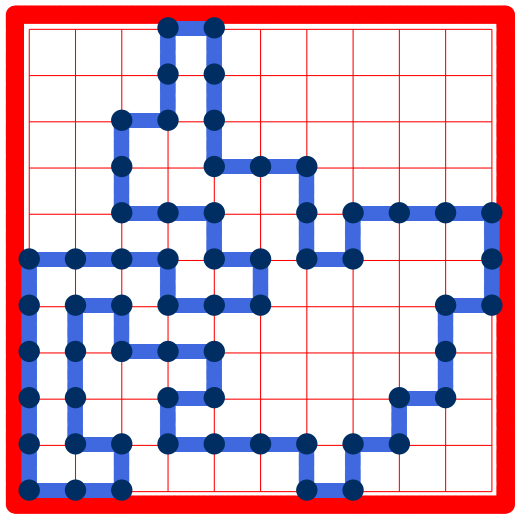

\centering
\input Figures/figure01.tex
\caption{A lattice polygon confined to a square in $\IntZ^2$ is 
a two dimensional model of a ring polymer which is compressed in 
a cavity.  In this particular diagram the length of the side of the
confining square is $10$, and it contains $11^2=121$ lattice sites. We say that
the \textit{dimension} of the confining square is $11$ and its
\textit{volume} is $121$.}
\label{figure01} 
\end{figure}

Knotting in self-avoiding walk models of polymers has received much 
attention in the literature \cite{SW88,P89,JvRW90,OTJW96}.  The usual model
for knotting in ring polymers is a lattice polygon in a three dimensional
lattice, often called a \textit{lattice knot}. It is known that 
the conformational entropy of a lattice knot is a function of its 
knot type \cite{SW88,P89}.  That is, the conformational entropy of an 
unknotted cubic lattice polygon is different from that of a polygon of
unrestricted knot type \cite{SW88,P89}; see references \cite{JvRW90,WJvR93}
for numerical evidence.   The entropy of lattice knots continues to be
the subject of numerous studies, including more recent work in
references \cite{MOSZ04,BOA10}.  In references \cite{MMOS06,MLY11,MO12} 
the effects of a confining space on the entropy of a model
of a ring polymer of fixed knot type was examined.

In this paper the entropic properties of a self-avoiding walk model of a 
confined ring polymer of fixed knot type are examined. The ring polymer is 
modelled by a closed self-avoiding walk (called a \textit{lattice polygon}) which is
confined to a cube in the cubic lattice $\IntZ^3$ (see figure \ref{figure01} for 
the two dimensional analog of this model).  A polygon confined to a cube 
will be called a \textit{compressed polygon}, and if its knot type is
fixed, then it is a \textit{compressed lattice knot}.

Compressed lattice polygons are models of biopolymers in confined spaces
(such as in living cells) \cite{D62,deG79}.  Other models of compressed 
polygons were considered in references \cite{LC08,LC12}, as a coarse
grained lattice model, and using a bond fluctuation model in reference \cite{CK88}.

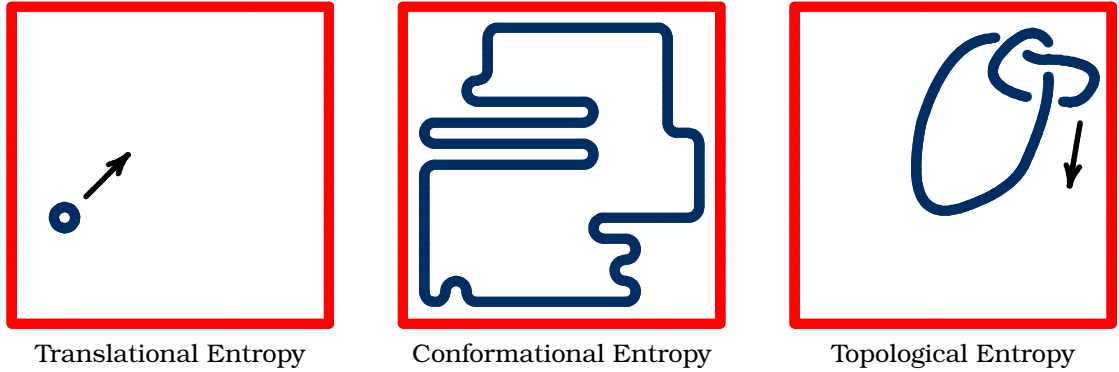
\begin{figure}[t]
\hspace{.5cm}
\scalebox{0.8}{
\begin{minipage}[t]{0.25\textwidth}
\input{Figures/figure02-1.tex}
\end{minipage} }
\hspace{0.5cm}
\scalebox{0.8}{
\begin{minipage}[t]{0.25\textwidth}
\input{Figures/figure02-2.tex}
\end{minipage} }
\hspace{0.5cm}
\scalebox{0.8}{
\begin{minipage}[t]{0.25\textwidth}
\input{Figures/figure02-3.tex}
\end{minipage} }
\caption{The entropy of a confined ring polymer of length $n$ in a cube 
of dimension $L$ in $d$ dimensions.  If the polymer is very short then it gains entropy from 
translational degrees of freedom  (left panel).  If it is long, then its entropy is primarily
determined by conformational degrees of freedom (middle panel). The crossover 
between the translational and conformational regimes should occur 
when the linear extent of the polymer approaches $L$. Since the linear
extent of an unconfined polygon is $O(n^\nu)$, this crossover occurs when
$O(n^\nu) \sim L$, where $\nu$ is the metric exponent of the polymer.
In a good solvent the Flory values of $\nu$ are $\nu=\frac{3}{4}$ if
$d=2$, and $\nu=\frac{3}{5}$ if $d=3$.    The concentration
of monomers at the crossover is $O(n/V) = O(L^{1/\nu - d})$ since the volume
of the confining cube is $V=L^d$.  That is, in $d=2$ the crossover
concentration is $\phi_a = O(L^{-2/3})$, and, in $d=3$, $\phi_a = O(L^{-4/3})$.
With increasing $L$ the crossover occurs at decreasing concentration.
If $d=3$ then the polygon may be knotted, and there are topological degrees
of freedom contributing to the free energy (right panel).}
\label{figure02}  
\end{figure}

In this paper the following questions are examined:
(1) How are the translational and configurational free energy of the 
polygon dependent on the concentration of monomers (modelled by 
vertices in the polygon) in the confining cube, and 
(2) what is the impact of topology (that is, of knot type) on the free energy of the polymer?

With respect to the first question, the model is a lattice knot confined
to a cube in $\IntZ^3$ (see reference \cite{GJvR18} for a model of a self-avoiding walk
confined to a square in the square lattice).  The confining cube has side-length
$L{-}1$ and contains $L^3$ lattice sites -- in this case the \textit{dimension} of
the cube is said to be $L$ and its \textit{volume} is $V=L^3$.  The concentration
$\phi$ of monomers in the lattice polygon is the number of monomers (vertices) per unit volume.
The concentration of solvent molecules (lattice vertices in the cube disjoint with
the lattice polygon) is $\phi_s = 1-\phi$.  It is assumed that the monomers in the
polygon can freely occupy lattice sites, and there is no interaction between
monomers and lattice sites (solvent molecules).  Monomer-monomer interaction
is a hard core repulsion due to self-avoidance of the polygon.

There are both translational and conformational contributions to the entropy
in the model, as shown in figure \ref{figure02}.  At low concentrations the polygon is 
very short, and it has low conformational entropy but can be translated inside 
the (relatively large) confining space (see the left panel in figure \ref{figure02}).  
That is, there are translation degrees of freedom which makes the dominant 
contribution to the entropy.  For a polygon of length $n$ in a cube of dimensions 
$L$ this is the low concentration regime, and it is encountered when $n \ll L$.
As shown in figure \ref{figure02}, the low concentration regime has
$\phi < \phi_a = O(L^{1/\nu - d})$. 

At high concentration the polygon has no translational 
degrees of freedom, but a large number of conformations;  see the middle
panel of figure \ref{figure02}.  In this case the dominant contribution
to the entropy is due to conformational degrees of freedom.
Thermodynamic quantities (for example, the osmotic pressure of monomers
in the lattice polygon, or of solvent molecules) can be determined from 
the free energy, and these are functions of the concentration $\phi$. 
In both the low and high concentration regimes the free energy will be a
function of knot type, as shown in the right panel in figure \ref{figure02}.

\begin{figure}[t]
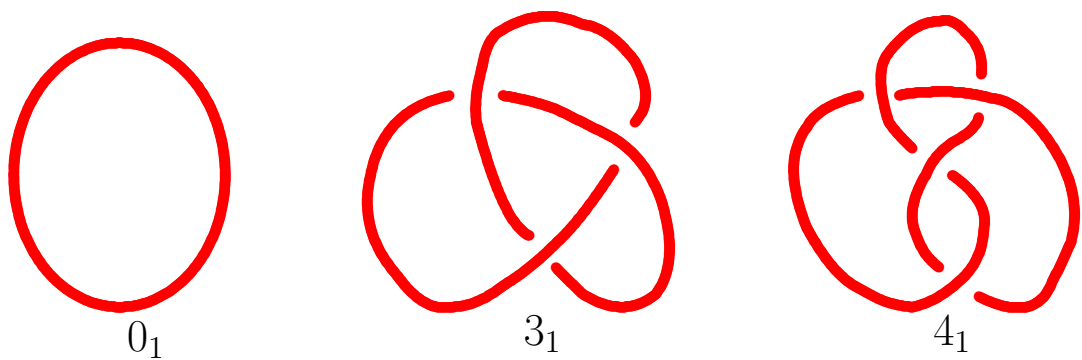

\input Figures/figure03.tex
\caption{The unknot ($0_1$), the trefoil ($3_1$) and the figure eight ($4_1$).}
\label{figure03}  
\end{figure}

In section \ref{section3} the free energy of compressed lattice knots is
estimated numerically and the dependence of the free energy on the 
concentration of monomers and on knot type is examined
for the knot types the \textit{unknot} $0_1$, the \textit{trefoil knot} $3_1$, 
and the  \textit{figure eight knot} $4_1$ (see figure \ref{figure03}).  The
discussion will be organised using Flory-Huggins theory \cite{H42,Flory42,Flory,deG79}
and by estimating the Flory Interaction Parameter of lattice knots.

In section \ref{section2} a short review of Flory-Huggins theory is given, with 
the emphasis on making a connection between the free energy per unit volume
of a compressed lattice knot, and the \textit{free energy of mixing} in
Flory-Huggins theory (the free energy of mixing is obtained by including energy
terms due to monomer-solvent interactions, and discarding all terms which
are constant, or linear in the concentration, from the free energy).  The Flory-Huggins
free energy of mixing is also not a function of the coordination number
of the lattice.  Since the estimated free energy is a function of the coordination
number of the lattice, it is shown that introduction of a term linear in the
concentration in the free energy of mixing accounts for this 
dependence, giving a suitable model for analysing the free energy 
of compressed cubic lattice polygons.

Numerical data is presented in section \ref{section3}, and by 
using Flory-Huggins theory to model the free energy of compressed lattice
knots, the Flory Interaction Parameter $\chi$ for confined lattice
knots is estimated:
\begin{equation}
\chi = 0.18\pm 0.03.
\end{equation}
This shows that dependence of $\chi$ on knot type is small, and cannot be
resolved by the current data.  In the broader context, this numerical 
result also shows that $\chi$ is not equal to zero,
and so if the lattice is viewed as a ``solvent" in which the lattice knot is dissolved,
then it is a good solvent (and the conditions are not athermal).  The self-avoiding
constraint on the lattice knot is a short ranged repulsive force between monomers,
and the effect of this is to give a non-zero value to $\chi$.  The Flory interaction
parameter is also less than $\sfrac{1}{2}$, which means that the compressed
lattice knot is well away from $\theta$-conditions.

In section \ref{section4} the paper is concluded with a few final remarks.

\section{Flory-Huggins theory and the free energy of compressed lattice knots}
\label{section2}

The number of states of polygons of knot type $K$ and length $n$ in $\IntZ^3$
is denoted by $p_n(K)$.  The \textit{growth constant} $\mu_K$ of lattice
knots of knot type $K$ is given by the limiting
value of the logarithm of $p_n(K)$ per unit length (per edge or per step):
\begin{equation}
\log \mu_K =  \limsup_{n\to\infty} \sfrac{1}{n} \log p_n(K) 
\label{eqn2}
\end{equation}
where the limsup is taken through even values of $n$.  
If $K$ is the unknot, then it is known that this limit exists \cite{SW88}, and
it is denoted by $\log \mu_{01}$.  

If a lattice knot is compressed in a cavity, 
then its free energy is a function of the dimension and shape of the cavity,
and it also becomes a function of the concentration $\phi$ (of monomers per unit
volume).  This can be modelled by using Flory-Huggins theory \cite{H42,Flory} 
(this is a mean field theory for modelling the free energy of concentrated 
polymers solutions or polymer melts).  

Flory-Huggins theory is based on (1) a mean field estimate of the entropy, 
and (2) a second order approximation of monomer-monomer and 
monomer-solvent interactions.  The theory is concerned with the 
\textit{free energy of mixing}, rather than the total free energy, and so 
ignores linear and background terms contributing to the total free energy 
(it is, for example, not a function of the lattice coordination number $\gamma$
while the free energy of a lattice polymer is a function of $\gamma$).

If a single chain is considered, then the entropy per unit volume $V$, 
of a polymer in a confining space, is estimated by
\begin{equation}
- S_{site} (\phi)= \Sfrac{\phi}{n} \log \Sfrac{\phi}{n} + (1-\phi)\log(1-\phi)
\label{eqn1}  
\end{equation}
where $\phi = \sfrac{n}{V}$ is the volume fraction (or concentration) of monomers 
in a chain of length $n$ (degree of polymerization)
\cite{H42}.    

The entropy of mixing \cite{Flory42} is the difference between 
$S_{site}(\phi)$ and the weighted average of the entropy of pure solvent 
$S_{site}(0)$, and pure polymer $S_{site}(1)$:
\begin{equation}
\fl \hspace{0cm}
- S_{mix} = - S_{site}(\phi) + (1-\phi) S_{site}(0) + \phi\, S_{site}(1)
= \Sfrac{\phi}{n} \log \phi + (1-\phi)\log (1-\phi) .
\end{equation}
This cancels terms linear in $\phi$ and is independent of the coordination number
of the lattice. 

The energy of mixing per site (or per unit volume) has contributions 
from solvent-solvent,  monomer-monomer and monomer-solvent interactions.  
For example,  the contribution of monomer-solvent interactions is approximated by
$E_{MS} = T\, \chi_{MS}\, \phi\,(1{-}\phi)$, and is a \textit{two-body}
approximation, leaving out higher order contributions.  
Contributions from monomer-monomer and
solvent-solvent interactions are similarly given by
$E_{SS} = T\, \chi_{SS}\, (1{-}\phi)^2$ and
$E_{MM} = T\, \chi_{MM}\, \phi^2$.  Collecting the second order
contributions, and then computing the energy of mixing similar to
$S_{mix}$ above, gives the energy of mixing
\begin{equation}
E_{mix} = T\, \chi\, \phi(1-\phi).
\label{eqn3}   
\end{equation}
where $\chi = \chi_{MS} - \sfrac{1}{2}(\chi_{MM}+\chi_{SS})$ is the
\textit{Flory Interaction Parameter}.

The \textit{mean field free energy of mixing per site} $F_{mix}$ is given
by adding to ${-}S_{mix}$ the energy of mixing $\sfrac{1}{T}E_{mix}$ 
per site \cite{deG79}:
\begin{equation}
\fl \hspace{0.5cm}
\Sfrac{1}{T}\, F_{mix} = \Sfrac{1}{T} \, E_{mix} - S_{mix}
= \Sfrac{\phi}{n} \log \phi + (1-\phi)\log(1-\phi) + \chi\, \phi(1-\phi) ,
\label{eqn4}   
\end{equation}
Expanding for small values of $\phi$ gives
\begin{equation}
\Sfrac{1}{T}\, F_{mix} = \Sfrac{\phi}{n} \log \phi + \sfrac{1}{2} (1-2\chi)\, \phi^2 
+ \sfrac{1}{6} \phi^3 + \cdots .
\label{eqn5}  
\end{equation}
The (Edwards) \textit{excluded volume parameter} is the coefficient of the
quadratic term given by $\upsilon = 1-2\chi$ \cite{Edwards76}.  If $\upsilon=0$
then the polymer is in $\theta$-conditions and the first correction 
to $\sfrac{\phi}{n} \log \phi$ in $F_{mix}$ is the third order term in $\phi$.
If $\upsilon<0$ then the polymer is in a poor solvent, and if $\upsilon>0$ it is
in a good solvent.  Notice that there are no linear terms  in $\phi$ in the expansion.

The parameter $\chi$ is a measure of the (repulsive) interactions between solvent 
molecules and monomers.   It tends to be positive \cite{deG79} and increasing 
with $T$.   Good solvents have small values of $\chi$ and if $\chi=0$ then the solution 
is said to be ``athermal".   When $\chi=\sfrac{1}{2}$ then the solvent is marginal and the polymer
is in $\theta$-conditions \cite{deG79}.  For $\chi>\sfrac{1}{2}$ the solvent is poor and 
the polymer may collapse from a coil to a globule phase \cite{deG75,S75}.

In order to apply these ideas in figure \ref{figure01}, put $T=1$ and work in 
lattice units.  The derivation of $F_{mix}$ discards terms linear in $\phi$, and also does not
include multi-body interactions between monomers and solvent molecules.  Instead, it
only includes a two-body interaction between monomers and solvent molecules, and
introduces the term $\chi \, \phi(1{-}\phi)$ to account for this.   The total free energy
of the model in figure \ref{figure01} is dependent on the lattice coordination number,
and it will be shown that the addition of a term linear in $\phi$ can account for
this.

\begin{figure}
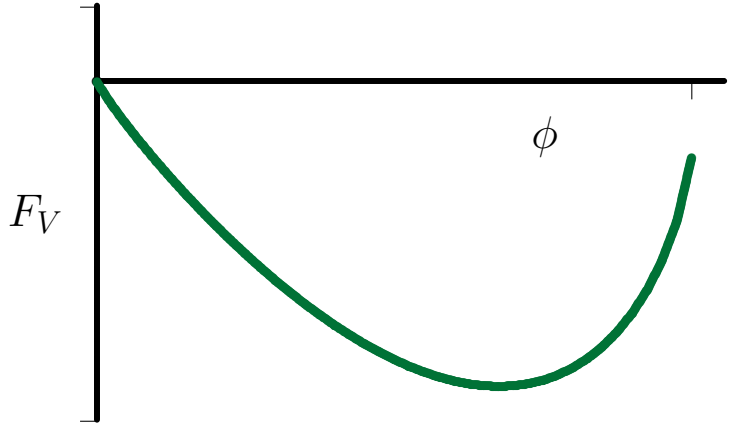

\input Figures/figure04.tex
\caption{The Flory-Huggins total free energy per lattice site $F_V$ (solid
curve) and the osmotic pressure (dashed curve) which
are obtained from the free energy of mixing $F_{mix}$
modified by substracting from it the linear term $0.10\, \phi$.
In this schematic drawing, $N=10$ and $\chi=\frac{1}{4}$.}
\label{figure04}  
\end{figure}

The total free energy of the model in figure \ref{figure01} is given by $F_{tot} 
= -\log p_{n,L} (K)$, where $p_{n,L} (K)$ is the total number of conformations of a
lattice polygon of knot type $K$ and with placements in the containing volume 
$V$ of dimensions $L^3$ counted as distinct.  The free energy per unit volume 
(or per lattice site) is given by
\begin{equation}
\hspace{-1cm}
F_V (K) = - \Sfrac{1}{V} \log p_{n,L} (K) .
\label{eqn8}
\end{equation}
The Flory free energy of mixing $F_{mix}$ in equation \Ref{eqn1} is also a free
energy per unit volume, and so is related to $F_V$.  However, $F_V$ is a function
of the lattice coordination number, and linear terms were discarded
in the mean field derivation of $F_{mix}$.  Thus,  the free energy of mixing $F_{mix}$
is a mean field approximation to $F_V$ up to missing linear terms in 
$\phi$ (and also up to cubic and higher order terms in $\phi$), and these
terms should be a function of the lattice coordination number.

That is, a mean field assumption for $F_V$, based on the Flory-Huggins free energy of
mixing, would be
\begin{equation}
F_V = a_0\,\phi + \Sfrac{\phi}{n} \log \phi + (1-\phi) \log (1-\phi)  - \chi\, \phi^2,
\label{eqn3A}   
\end{equation}
where $a_0$ is a constant dependent on the lattice coordination number, as
will be seen below.

In figure \ref{figure04} a schematic diagram for $F_V$ with $\chi=\sfrac{1}{4}$
and $a_0 = -0.10$ is shown.  $F_V$ is represented by the solid curve and it is a convex
function of $\phi$.  

The \textit{free energy per monomer} of the lattice knot is given by
\begin{equation}
f_t (\phi) = - \sfrac{1}{n} \log p_{n,L} (K) = \sfrac{1}{\phi}\, F_V.
\label{eqn10}   
\end{equation}
Using the model in equation \Ref{eqn19A},  the mean field Flory-Huggins
expression for $f_t(\phi)$ is 
\begin{equation}
\hspace{-1cm}
f_t(\phi) 
= a_0 + \Sfrac{1}{V\phi} \log \phi + \Sfrac{1-\phi}{\phi} \log(1-\phi) - \chi_{L,K}\,\phi ,
\end{equation}
where the possibility that $\chi_{L,K}$ is a function of $L$ is explicitly indicated.
Taking $L\to\infty$ (and so $V\to\infty$) gives the limiting curve
\begin{equation}
\lim_{L\to\infty} f_t(\phi) 
= a_0 + \Sfrac{1-\phi}{\phi} \log(1-\phi) - \chi_{K}\, \phi,
\label{eqn24}   
\end{equation}
provided that $\phi>0$ and where $\chi_{K}$ is the limiting value of the
Flory interaction parameter for these models of lattice knots.    
Taking the limit $\phi\to 0^+$ on the right hand side gives
\begin{equation}
\lim_{\phi\to 0^+} \lim_{L\to\infty} f_t(\phi) = a_0 - 1.
\end{equation}
This is the zero concentration limit, and this should be equal to 
$-\log \mu_{K}$ (where $\mu_{K}$ is the growth constant of lattice
knots of type $K$ since in the dilute limit the polygon is small compared to the
confining cube and so has self-avoiding polygon statistics -- see equation
\Ref{eqn2}).   The growth constant $\mu_{K}$ is dependent on the 
coordination number of the lattice and this shows that for the cubic lattice
knot there, there is a contribution of $n \log \mu_{K} + o(n)$ to the total
free energy $F_{tot}$.  Dividing by $V$, and then by $\phi$, as in equations
\Ref{eqn3A} and \Ref{eqn10}  gives the contribution $a_0-1 = \log \mu_{K} + o(1)$.

To see this, place a polygon of length $L$ 
and knot type $K$ in a cube of dimensions $L^3$.    The number
of polygons in the cube is $O(L^3)\times p_L(K)$ (where $p_L(K)$ is the number of 
polygons of length $L$ and knot type $K$).  Taking logarithms, dividing by
$L$ and taking the limit superior as $L\to\infty$ gives $- \log  \mu_{K}$, 
while the concentration goes to zero as $O(L^{-2})$.  This shows that
$a_0 = 1 - \log  \mu_{K}$.  That is, fitting numerical data against $f_t(\phi)$ 
for finite values of $L$ should give estimates of $a_0$ which approaches 
$1-\log  \mu_{K}$ as $L$ increases, while $\chi_L$ approaches a limiting 
value which is equal to the Flory Interaction Parameter of the polygons.

\begin{table}
\begin{center}
\begin{tabular}{| c | c c c | }
\hline
$L^3$  &  $0_1$ ($n=4$) &  $3_1$ ($n=24$) &  $4_1$ ($n=30$) \cr
\hline
$2^3$ & $6$ & ${-}$ & ${-}$  \cr
$3^3$ & $36$ & ${-}$ & ${-}$  \cr
$4^3$ & $108$ & $4168^{*}$ & $864^{*}$  \cr
$5^3$ & $240$ & $30104$ & $18048$  \cr
$6^3$ & $450$ & $97752$ & $73440$  \cr
$7^3$ & $756$ & $227080$ & $188928$  \cr
$8^3$ & $1176$ & $438056$ & $386400$  \cr
$9^3$ & $1728$ & $750648$ & $687744$    \cr
$10^3$ & $2430$ & $1184824$ & $1114848$ \cr
$11^3$ & $3300$ & $1760552$ & $1689600$  \cr
$12^3$ & $4356$ & $2497800$ & $2433888$ \cr
$13^3$ & $5616$ & $3416536$ & $3369600$ \cr
$14^3$ & $7098$ & $4536728$ & $4518624$ \cr
$15^3$ & $8820$ & $5878344$ & $5902848$ \cr
\hline
\end{tabular}
\caption{Counts of minimal length polygons in $L^3$}
\label{Counts}
\end{center}
\end{table}

\section{Numerical data}
\label{section3}

The GAS algorithm \cite{JvRR09,JvRR11} was implemented with BFACF elementary
moves \cite{BF81,ACF83,JvRW91} to sample lattice knots 
in $\IntZ^3$ along a sequence in state space \cite{JvRR11}.   
Confining the lattice knot to a cube reduces the size of state space, and
creates lattice knot irreducibility classes fitting inside the cube.  This reduction
in the size of the irreducibility class is a model of the reduction in entropy
when a ring polymer is confined to a cavity.  In each model (defined by
the knot type of the compressed lattice knots, and by the dimension
of the confining cube) there is an irreducibility class which contains
the minimum length lattice knots.

In the case of the unknot the minimal length is $4$ and there are three minimal 
length realisations in the cubic lattice, namely the unit square in each of the three lattice 
planes.  When confined in a cube of dimension $L$ (and side length $L{-}1$) 
these minimal length unknotted polygons are members 
of an irreducibility class which will be called the \textit{natural class} of 
compressed (unknotted) polygons.  That is, the natural class of lattice knots
in a cube of dimension $L$ is the set of all lattice knots in the cube
which can be reached from minimal length lattice knots in the cube.

The number of states of minimal length in the natural class of unknotted 
lattice knots can be counted, and these numbers are shown in the second 
column of table  \ref{Counts}.  For a cube of dimension $L$  the number of 
minimal length lattice knots of length $4$ is $3L(L{-}1)^2$.  For example, if 
$L=2$ (the cube with $8$ sites) there are $6$ polygons of length $4$.

\begin{figure}[t]
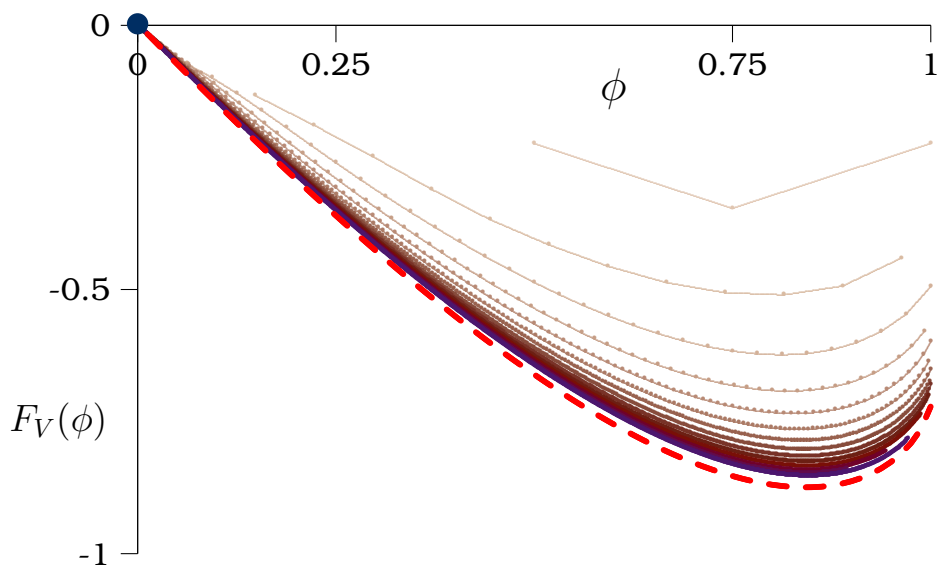

\input Figures/figure05.tex
\caption{The free energies $F_V(\phi)$ of compressed unknotted 
lattice knots confined to cubes of dimensions $L$, with 
$L\in\{3,4,5,\ldots,12\}$.  With increasing values of $L$ the curves
systematically decrease to an apparent limiting curve.  The
dashed line is the limiting curve given by Flory-Huggins theory
(see equation \Ref{eqn19A}).}
\label{figure05}  
\end{figure}

The situation is more complicated if the polygons are knotted.  
For example, for lattice knots of knot type $3_1$ (the trefoil) in a cube of dimension
$5$ (with $5^3$ sites), a computer count gives $30104$ lattice knots
of minimal length $n=24$ that can be placed inside the cube (where 
lattice knots equivalent under translations inside the cube are counted as 
distinct conformations).  These minimal length lattice knots are members of 
the natural class of compressed lattice knots of knot type $3_1$ (the trefoil) 
in the cube of dimension $5$ and with $125$ lattice sites.  

More generally, for the knot type $3_1$, $p_{24}(3_1)= 3328$, since there are
$3328$ distinct (up to translation) realisations of these minimal length lattice 
knots of length ($n=24$) \cite{SIA09}.  One-half of these will be left handed
trefoils, and the other half are right handed.  In a cube of dimension $3$ 
(with $3^3=27$ sites), none of these minimal lattice knots will fit, and so the natural 
class is empty.  In a cube of dimension $4$ (with $4^3=64$ sites) there are $3304$ 
minimal length lattice knots (of both left-handed and right-handed chirality) which 
can be placed inside the cube.  This gives a natural class containing $4168$ minimal length
lattice knots (with placements equivalent under translations counted as distinct).  
There are also $24$ minimal length lattice knots of type $3_1$ which will not 
fit into the cube of dimensions $4$.   In other words, the natural class of minimal 
length lattice knots does not include representatives of these polygons, and this 
is indicated by the $^{*}$ in table \ref{Counts}.  The GAS algorithm finds all these 
$4168$ minimal length states in the natural class (provided that rigid rotations and reflections 
of the entire polygon is added to the BFACF moves).

All minimal length lattice knots of type $3_1$ can be placed inside a cube
of dimension $5$, and the natural class contains $30104$ states.  The algorithm, 
if implemented with rigid rotations and reflections of the confined lattice knot 
inside the cube, finds all these states, so that all minimal length lattice knots in a
cube of dimensions $5$ are in the irreducibility class of the algorithm.  This is 
similarly true for larger cubes -- the number of minimal length states in 
each case is listed in table \ref{Counts}.

Similarly, there are $3648$ minimal length lattice knots of type $4_1$ (the minimal
length is $30$) \cite{SIA09}.  None of these minimal length lattice knots
can be placed in a cube of dimension $3$, but a total of $864$ lattice knots of
length $30$ and type $4_1$ can be placed in a cube of dimension $4$.
The GAS algorithm, implemented with rigid rotations and reflections,
detects all $864$ of these minimal  length states in a cube of dimension $4$, and so the
natural class for lattice knots of type $4_1$ in a cube of dimension
$4$ is defined in this way.   If the cube has dimension $5$, then the algorithm
detects all $3648$ minimal length lattice knots of type $4_1$, and they can
be placed $18048$ distinct ways, as shown in table \ref{Counts}.  

These observations are similarly true for larger cubes and the number 
of states of minimal length in the natural class is listed in table \ref{Counts}
and are detected by the algorithm.

\begin{table}[t!]
\begin{center}
\begin{tabular}{|| c | c c c || }
\hline  
\hline
$L$  &  $\chi_{01}$ &  $\chi_{31}$ &  $\chi_{41}$ \cr
\hline
3 & $0.4372$ & ${-}$ & ${-}$  \cr
4 & $0.3611$ & $0.9796$ & $1.2820$  \cr
5 & $0.3084$ & $0.5983$ & $0.7395$  \cr
6 & $0.2810$ & $0.4185$ & $0.4913$  \cr
7 & $0.2617$ & $0.3333$ & $0.3737$  \cr
8 & $0.2514$ & $0.2905$ & $0.3189$    \cr
9 & $0.2437$ & $0.2662$ & $0.2814$ \cr
10 & $0.2388$ & $0.2524$ & $0.2619$  \cr
11 & $0.2335$ & $0.2433$ & $0.2501$ \cr
12 & $0.2284$ & $0.2367$ & $0.2396$ \cr
13 & $0.2188$ & $0.2235$ & $0.2272$ \cr
14 & $0.2122$ & $0.2165$ & $0.2203$ \cr
15 & $0.2176$ & $0.2246$ & $0.2239$ \cr
\hline\hline
\end{tabular}
\caption{Estimated Flory interaction parameters from $F_V(\phi)$.}
\label{tablechi}   
\end{center}
\end{table}

The GAS algorithm was implemented using the data in table \ref{Counts} and 
it sampled along GAS sequences in parallel on a multiprocessor computer.  A 
total of $400$ sequences were realised, each of length $2L\times 10^6$ iterations.
For example, for $L=10$, a total of $2\times 10^7$ lattice knots in a 
cube of dimension $10$ were sampled along each sequence,
for a total of $8\times 10^9$ iterations in total.  The resulting data are
approximate counts of lattice knots of length $n$ in the natural class
in a cube of dimension $L$;  these are estimates of $p_{n,L}(K)$, the number
of lattice knots in the natural class, of knot type $K$, and length $n$.

The free energy per unit volume of compressed lattice knots of type
$K$ is given in equation \Ref{eqn8}, and in figure \ref{figure05} 
the estimates of the free energy per lattice site $F_V(0_1)$ of compressed 
unknotted lattice knots in cubes of volume $V=L^d$  
are plotted as a function of the concentration of vertices 
$\phi = \sfrac{n}{V}$.  The general appearance 
of these curves is similar to the curve due to Flory-Huggins theory 
shown in figure \ref{figure04}.

\begin{figure}[t]
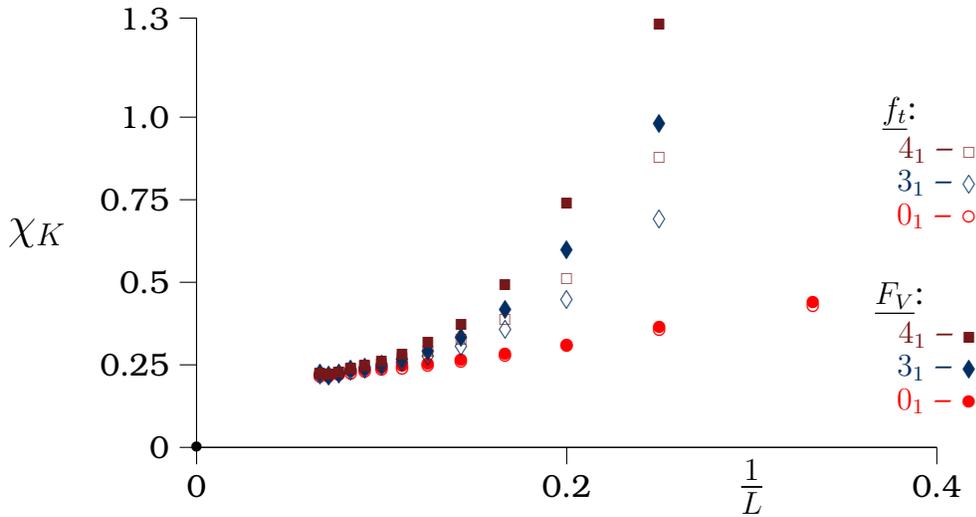

\input Figures/figure06.tex
\caption{Estimates of the Flory Interaction Parameter $\chi_K$ for knot types
the unknot ({\large$\bullet$}), the trefoil ($\blacklozenge$), and the figure eight knot
({\scriptsize$\blacksquare$}).  The solid points are estimates
determined from the free energy $F_V$,  while open points are estimates
determined from $f_t$.  The data are plotted against $1/L$ and clearly approaches
a limiting value as $L$ increases.}
\label{figure06}  
\end{figure}

\subsection{Unknotted compressed lattice knots}

The Flory-Huggins mean field expression for $F_V(\phi)$ is given by equation
\Ref{eqn3A}.  Since $\phi=\sfrac{n}{V}$, the contribution of the second term to
this expression is small for larger values of $\phi$, and the approximation
\begin{equation}
F_V(0_1) \simeq a_0\, \phi + (1-\phi)\log(1-\phi) - \chi_{01} \, \phi^2
\label{eqn15}   
\end{equation}
may be used to model the data in figure \ref{figure05}.  For example, when $L=10$
then a least squares fit to the data gives
\begin{equation}
F_V(0_1)\vert_{L=10} \approx -0.4252\, \phi + (1-\phi)\log(1-\phi)  - 0.2388\, \phi^2 .
\end{equation}
This suggests that $a_0 = -0.4252\ldots$, and the estimate for $\chi_{01}$ is a 
finite $L$ value of the Flory-Huggins interaction parameter: 
$\chi_{01}\vert_{L=10} = 0.2388\ldots$.

The remaining estimates of $\chi_{01}$ for $3\leq L \leq 15$ are listed in the second
column of table \ref{tablechi}.  These data are plotted in figure \ref{figure06} 
against $\sfrac{1}{n}$ (solid $\bullet$'s).  There are some curvature in the data 
for small values of $L$, but an extrapolation using a quadratic polynomial for 
$L\geq 7$ gives the limiting value $\chi_{01} = 0.18$,  while a linear extrapolation also 
gives $\chi_{01} = 0.18$. Including all the data for all values of $L$ instead gives $\chi_{01}=0.18$ 
(a quadratic extrapolation), and $\chi_{01}=0.15$ (a linear extrapolation).
Thus, we take as our best estimate $\chi_{01}=0.18\pm 0.03$.   

\begin{figure}[h!]
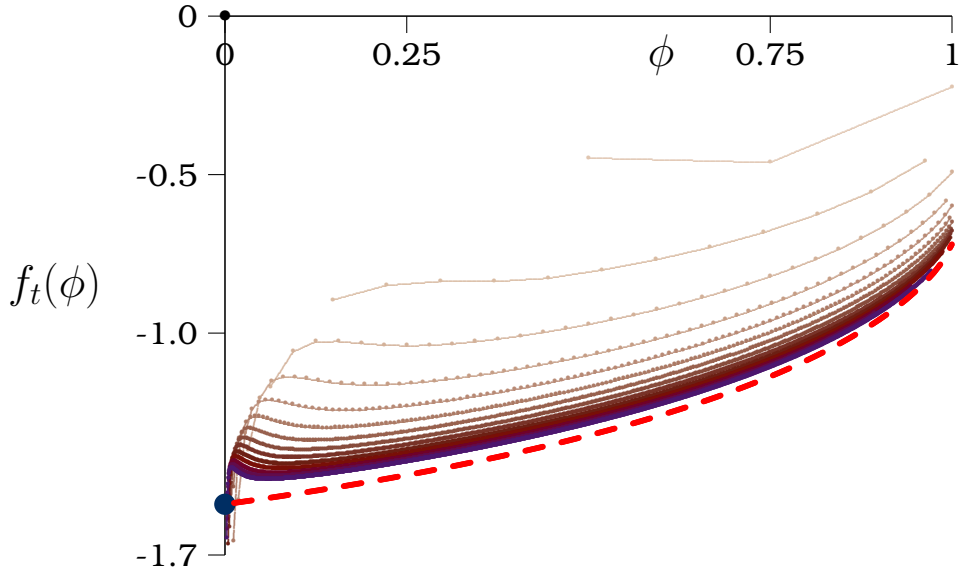

\input Figures/figure07.tex
\caption{The free energies $f_t(\phi)$ of compressed unknotted lattice 
polygons confined in cubes of dimensions $L^3$, with 
$L\in\{3,4,5,\ldots,15\}$.  With increasing values of $L$ the curves
systematically decrease to an apparent limiting curve.  Notice the behaviour
at small $\phi$ where there is a sharp decrease in $f_t(\phi)$ due to 
translational degrees of freedom at low concentrations.  The
dashed line is the limiting curve given by the approximation in
equation \Ref{eqn17} with $a_0 = 1 - \log \mu_{01}$ and
$\chi_{01} = 0.18$.}
\label{figure07}  
\end{figure}

The free energy per monomer $f_t(\phi)$ (see equation \Ref{eqn10}) is plotted in
figure \ref{figure07}.  The sharp decrease in $f_t(\phi)$ for increasing $\phi$ at low 
concentration is due a reduction in translational degrees of freedom with increasing
concentration when the lattice polygon is very short (and can explore the volume of 
the confining cube freely).  With increasing values of $L$ the free energy
$f_t(\phi)$ appears to approach a limiting curve.  There is a critical 
concentration $\phi^*$ where $f_t(\phi)$ has a local minimum.  With increasing 
$L$, the critical concentration $\phi^*$ appears to decrease towards zero.

\begin{table}
\begin{center}
\begin{tabular}{|| c | c c c || }
\hline\hline
$L$  &  $\chi_{01}$ &  $\chi_{31}$ &  $\chi_{41}$ \cr
\hline
3 & $0.4265$ & ${-}$ & ${-}$  \cr
4 & $0.3530$ & $0.6906$ & $0.8784$  \cr
5 & $0.3046$ & $0.4482$ & $0.5115$  \cr
6 & $0.2748$ & $0.3564$ & $0.3886$  \cr
7 & $0.2561$ & $0.3053$ & $0.3265$  \cr
8 & $0.2437$ & $0.2760$ & $0.2895$    \cr
9 & $0.2370$ & $0.2569$ & $0.2686$ \cr
10 & $0.2320$ & $0.2471$ & $0.2531$  \cr
11 & $0.2260$ & $0.2383$ & $0.2450$ \cr
12 & $0.2220$ & $0.2325$ & $0.2370$ \cr
13 & $0.2165$ & $0.2231$ & $0.2270$ \cr
14 & $0.2192$ & $0.2180$ & $0.2225$ \cr
15 & $0.2132$ & $0.2193$ & $0.2218$ \cr
\hline\hline
\end{tabular}
\caption{Estimated Flory interaction parameters from $f_t$}
\label{tablechift}   
\end{center}
\end{table}

When $\phi< \phi^*$ the lattice knot is short and translational degrees of freedom may 
make a contribution to the free energy.  As $L$ increases $\phi^*$ should decrease
to zero, and the minimum in the free energy should approach $-\log \mu_{01}$, 
where $\mu_{01}$ is the growth constant of unknotted lattice polygons
(see equation \Ref{eqn2}, and, for example, references \cite{JvRW90,WJvR93}).  This is seen
by noticing that at a concentration $\phi_L=\sfrac{L}{V}$ polygons have length $L$
and the total number of states is of order $O(L^3\, p_{L,L}) \simeq O(L^3\, p_L(0_1))$, where
$p_L(0_1)$ is the number of unknotted lattice polygons of length $L$.  Taking
logarithms, dividing by $L$, and taking the limit superior as $L\to\infty$, 
gives $\log \mu_{01}$,  while $\phi_L \to 0^+$.

The Flory-Huggins approximation to $f_t(\phi)$ is given by equation \Ref{eqn24}:
\begin{equation}
\hspace{-1cm}
f_t(\phi) \simeq a_0 + \Sfrac{1-\phi}{\phi}\, \log (1-\phi) - \chi_{01} \, \phi .
\label{eqn17}   
\end{equation}
Taking the derivative of the right hand side shows that this is an increasing function of $\phi$. 
However the curves in figure \ref{figure07} are not monotone, but go through a
local minimum at a critical concentration $\phi^*$ before increasing monotone
as $\phi$ approaches $1$. 

\begin{figure}[t!]
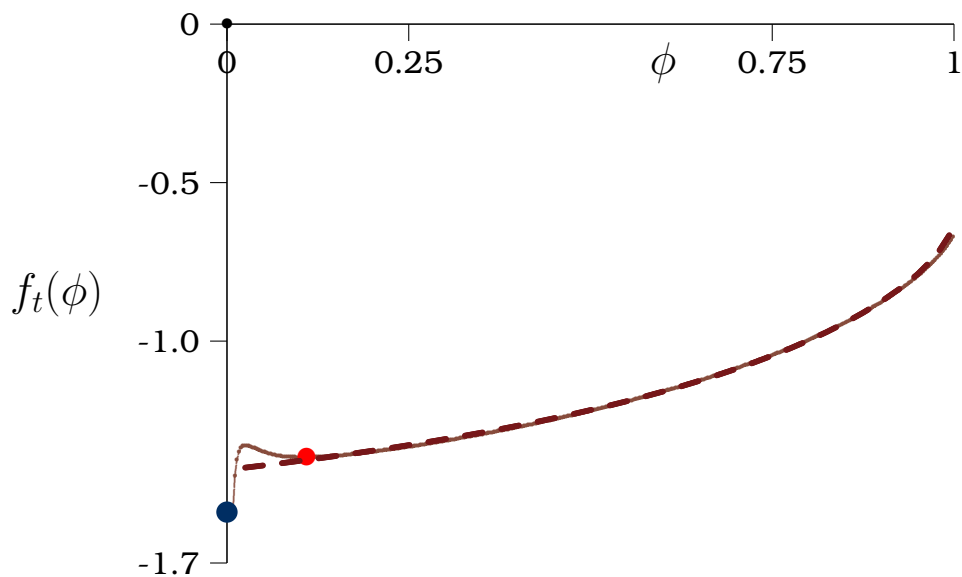

\input Figures/figure08.tex
\caption{The free energies $f_t(\phi)$ of compressed unknotted lattice 
polygons in a cube of dimension $9$.  The local minimum is indicated by
the bullet and it occurs at a concentration $\phi^*$.  If $L\to\infty$, then
$\phi^*$ should approach zero, and the local minimum in the free energy
should approach $- \log \mu_{01}$, where $\mu_{01}$ is the growth constant 
of lattice polygons of knot type $0_1$.  In this case (for $L=9$) the local 
minimum is at $-1.3697$, and the critical concentration is $\phi^* = 0.1097$.  
The dashed curve is the least squares fit of equation \Ref{eqn17} for 
$\phi>\phi^*$.}
\label{figure08}  
\end{figure}

In order to estimate the Flory Interaction
Parameter $\chi_{10}$ from these data, a least squares fit of equation \Ref{eqn17}
was done against the data for $\phi>\phi^*$ (that is, on numerical data for
concentrations where $f_t(\phi)$ is monotone increasing).  This is shown in figure
\ref{figure08} for $L=9$.  The location of the local minimum in $f_t(\phi)$ is indicated by a bullet.
It is expected that $\phi^* \to 0^+$ and thus $f_t (\phi^*) \to - \log \mu_{01}$ 
as $L \to\infty$.  This gives an estimate of the limiting value of the parameter $a_0$ in
equation \Ref{eqn17}, namely $a_0 \to 1 - \log \mu_{01}$ as $L\to\infty$.
The bullets on the vertical axes in figures \ref{figure07} and \ref{figure08} denote
$-\log \mu_{01}$.

The least squares estimates of $\chi_{01}$ from the data in figure \ref{figure07} are 
functions of $L$.  As $L$ increases these should give an extrapolated estimate of 
the Flory Interaction Parameter of unknotted lattice polygons.   
In a cube of dimension $9$ (as in figure \ref{figure08}) a least squares
fit gives $a_0\vert_{L=9} = -0.4107\ldots$ and $\chi_{01} \vert_{L=9} = 0.2370\ldots$.
The estimates of $\chi_{01}$ for the values of $L\in\{3,4,5,\ldots,15\}$ are listed in 
table \ref{tablechift}.  Extrapolating the data with a quadratic gives the limiting estimate
$\chi_{01}=0.17$, and a linear extrapolation of the data for $L\geq 9$ gives
$\chi_{01}=0.18$.  These results are consistent with those determined from $F_V$ 
above.

\begin{figure}[t]
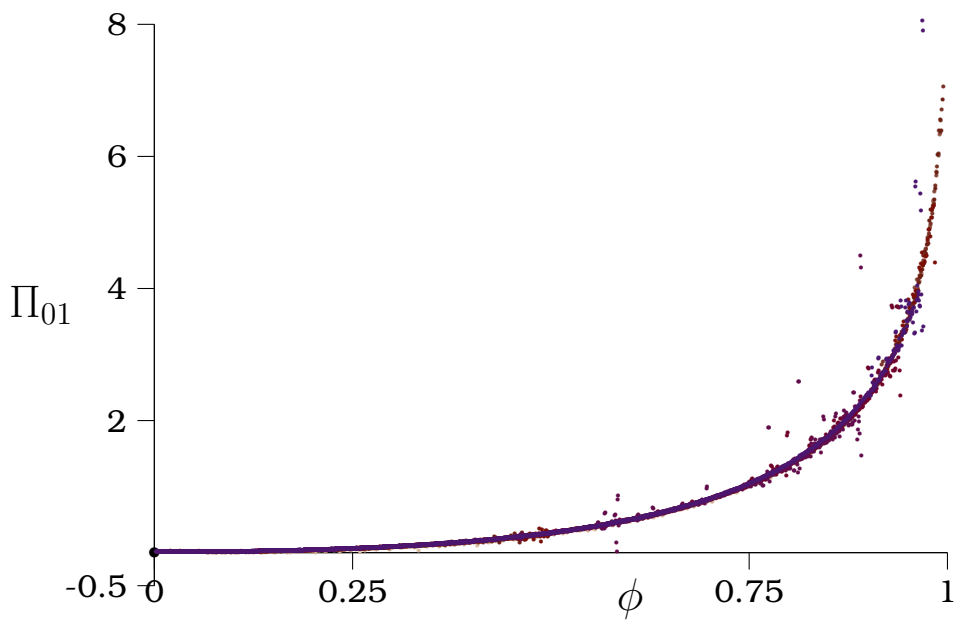

\input Figures/figure09.tex
\caption{Osmotic pressure $\Pi_{01}$ for compressed unknotted lattice
knots determined by equation \Ref{eqn19B}.}
\label{figure09}  
\end{figure}

\begin{figure}[t]
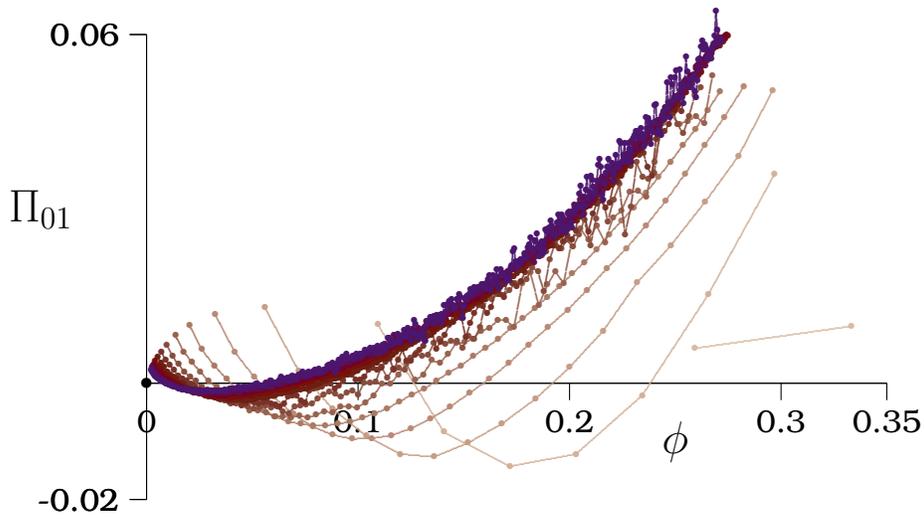

\input Figures/figure10.tex
\caption{The osmotic pressure $\Pi_{01}$ for compressed unknotted
lattice knots at low concentration.  The pressure turns negative for low values
of the concentration, and has a minimum at a critical concentration.}
\label{figure10}  
\end{figure}

Taken together, the consistent results for the Flory interaction parameter determined from
$F_V(0_1)$ and $f_t(0_1)$ indicates the best numerical estimate
of the limiting value of $\chi_{01}$ from the numerical data in this study:
\begin{equation}
\chi_{01} = 0.18 \pm 0.03 .
\label{eqn18A}   
\end{equation}
The value of $a_0$ in equation \Ref{eqn15} is given by $1-\log \mu_{01}$, and
since numerical results indicate that $\mu_{01} \approx \mu_3$ to at
least 5 signficant digits \cite{JvRW90}, one may approximate
$a_0 \approx 1 - \log \mu_3 = -0.5442\ldots$ (here $\mu_3$ is the self-avoiding 
walk growth constant in the cubic lattice). This gives the following Flory-Huggins
mean fied approximation to the free energy $F_V(0_1)$ of compressed 
unknotted polygons:
\begin{equation}
F_V(0_1) \approx -0.544\, \phi + (1-\phi)\log(1-\phi) - 0.18\, \phi^2 .
\label{eqn19A}   
\end{equation}
This is plotted as the dashed curve in figure \ref{figure05} and it is a good
approximation of $F_V(0_1)$ over the entire range of $\phi\in[0,1]$.
Since $f_t(0_1) = \sfrac{1}{\phi} F_V(0_1)$, the model above
is also a model of the free energy per unit length, and this is shown by
the dashed curve in figure \ref{figure07}.

From the free energy $f_t(\phi)$ one may compute other thermodynamic 
quantities, and in particular the osmotic pressure of monomers in the 
polymer (that is, the tendency of monomers to enter or exit the lattice polymer,
changing its length).  Taking the derivative of $f_t(\phi)$ to $\phi$ shows 
that the osmotic pressure of monomers in unknotted lattice knots is given by
\begin{equation}
\Pi_{01} = \phi^2 \Sfrac{d}{d\phi} \, f_t(\phi) .
\label{eqn19B}
\end{equation}
The data in figure \ref{figure07} is of sufficient high quality so that a 
numerical derivative can be taken to estimate $\Pi_{01}$.  Using a central
second order numerical approximation to the derivative gives the result
shown in figure \ref{figure09}.  The data collapse to a single curve for
large values of $\phi$, but closer examination for low concentration 
in figure \ref{figure10} show dependence on $L$ and negative osmotic pressures.

\subsection{Compressed lattice knots of knot types $3_1$ and $4_1$}

\begin{figure}[t]
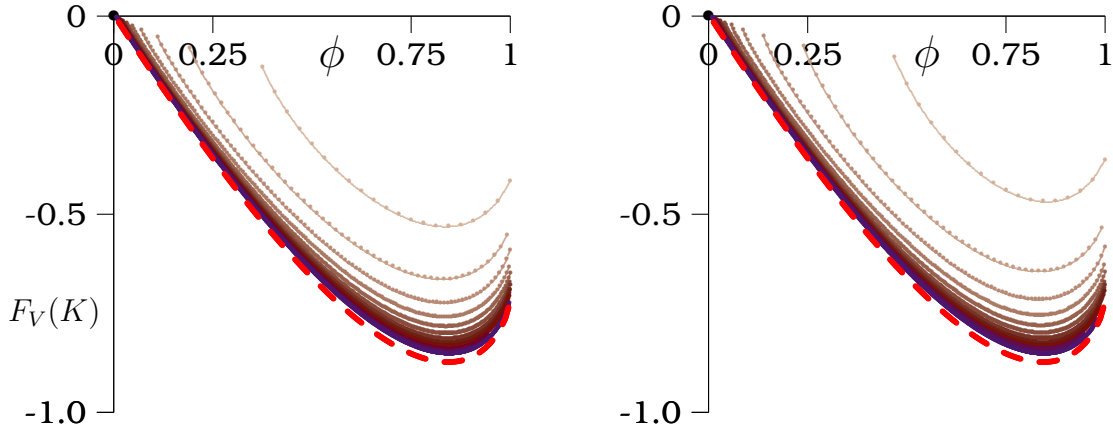

\input Figures/figure13.tex
\caption{The free energies $F_V(K)$ of compressed lattice knots confined to
cubes of dimension $L$, with $L\in\{4,5,\ldots,15\}$ and for knot types
the trefoil (left panel) and the figure eight knot (right panel).  
With increasing values of $L$ the curves systematically decrease to an apparent 
limiting free energy per unit volume given approximately by equation
\Ref{eqn22D}.  In this particular graphs, $\chi_K = 0.18$ and $a_0 = 
1-\log \mu_3(0_1)$; see equation \Ref{eqn19A}.  Notice the differences
between the left and right panels for smaller values of $L$.}
\label{figure13}  
\end{figure}

The free energy per unit volume $F_V(K)$ of lattice knots of knot types $3_1$ and $4_1$
are plotted in figure \ref{figure13} for lattice knots confined to cubes 
of dimension $L$ and for $L=4,5,\ldots,12$.  With increasing $L$ the curves 
appear to approach a limiting curve which looks similar to the curves for
$F_V(0_1)$ in figure \ref{figure05} (the free energy per unit volume for the unknot).

The model in equation \Ref{eqn3A} gives 
\begin{equation}
F_V (K) \simeq a_0\, \phi + (1-\phi)\log(1-\phi) - \chi_K \, \phi^2
\label{eqn22D}   
\end{equation}
where $\chi_K$ is the Flory Interaction Parameter for lattice knots of type $K$.  Least
squares fits of this model to the data in figure \ref{figure13} give the
estimates for  $\chi_{31}$ and $\chi_{41}$ in the third and fourth columns
of table \ref{tablechi}.  The estimates for $\chi_{31}$ and $\chi_{41}$
are significantly larger than for $\chi_{01}$ for small values of $L$ (say for $L<8$), 
but for $L>10$ the estimates converge to those measured for $\chi_{01}$.
These estimates are plotted in figure \ref{figure06}, with $\chi_{31}$ represented 
by \scalebox{0.8}{$\blacklozenge$}, and $\chi_{41}$ by \scalebox{0.7}{$\blacksquare$}.  
Clearly, with increasing $L$, these estimates appear to converge to the same limiting
values estimated for $\chi_{01}$ (denoted by $\bullet$).

\begin{figure}[t]
\input Figures/figure14.tex
\caption{The free energy $f_t(\phi)$ of compressed lattice knots of type
the trefoil ($3_1$) confined in cubes of dimension $L$, with 
$L\in\{4,5,\ldots,15\}$.  With increasing values of $L$ the curves
systematically decrease to a hypothetical limiting curve.  The
dashed line is given by equation \Ref{eqn17} with $\chi_{01}=0.18$ and
$a_0 = 1-\log \mu_3(0_1)$, showing that the values of the unknot
gives a good approximation to the limiting curve.}
\label{figure14}  
\end{figure}

\begin{figure}[t]
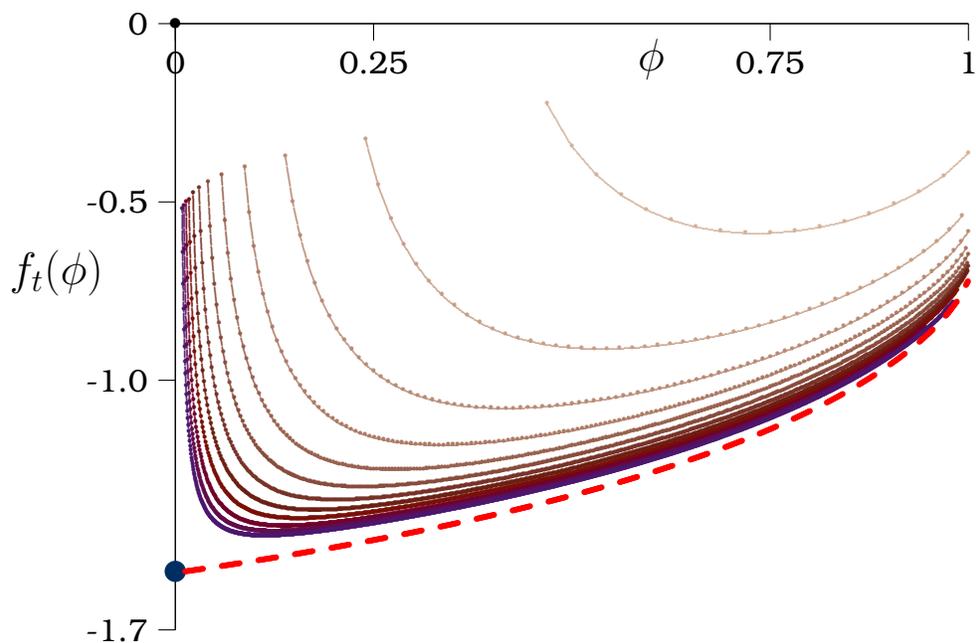

\input Figures/figure15.tex
\caption{The free energy $f_t(\phi)$ of compressed lattice knots of type
the figure eight knot ($4_1$) confined in cubes of dimension $L$, with 
$L\in\{4,5,\ldots,15\}$.  With increasing values of $L$ the curves
systematically decrease to a hypothetical limiting curve.  The
dashed line is given by equation \Ref{eqn17} with $\chi_{01}=0.18$ and
$a_0 = 1-\log \mu_3(0_1)$, showing that the values of the unknot
gives a good approximation to the limiting curve.}
\label{figure15}  
\end{figure}

\begin{figure}[t]
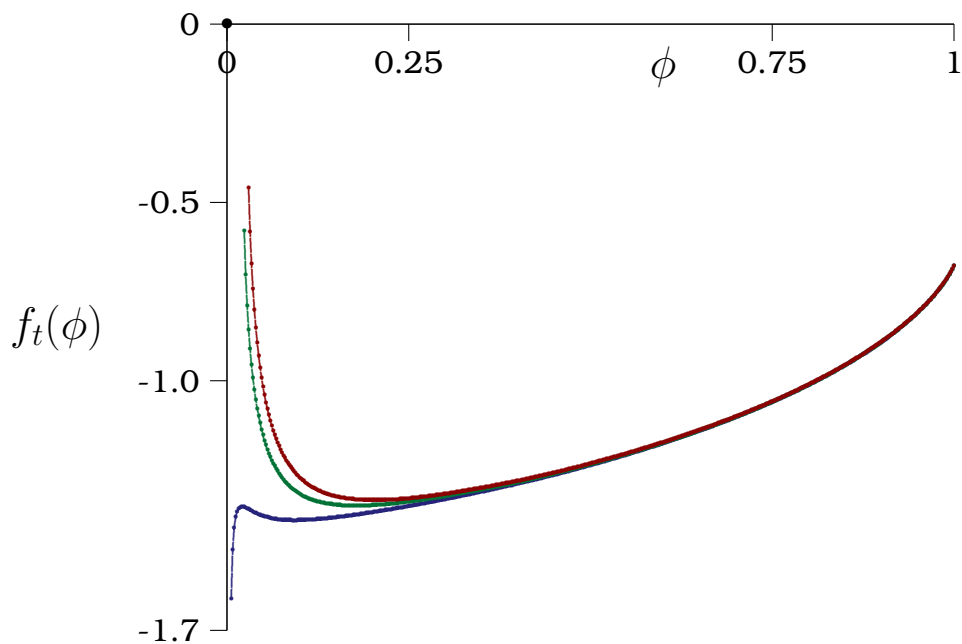

\input Figures/figure16.tex
\caption{The free energies $f_t(\phi)$ of compressed lattice knots with
$K=0_1$ (bottom curve -- the unknot), $K=3_1$ (middle curve -- the trefoil)
and $K=4_1$ (top curve -- the figure eight knot).  The lattice knots were
confined in a cube of dimension $L=10$.}
\label{figure16}  
\end{figure}

The free energies per unit length $f_t(\phi)$ for the trefoil knot and the
figure eight knot are shown in figures \ref{figure14} and \ref{figure15}
respectively.  These curves should be compared to the results for the
unknot in figure \ref{figure07}, which have substantial different behaviour.
These differences are also illustrated in figure \ref{figure16} for
$L=10$, and they are particularly prominent for low values of the concentration.
Minimal length lattice trefoil and figure eight knots have a substantial 
configurational entropy, in addition to contributions from translational degrees 
of freedom at low concentration.  On the other hand, minimal length lattice 
knots of knot type the unknot have little configurational entropy, 
but also has a relatively large contribution to its free energy due to translational 
degrees of freedom at low concentration.  These observations about contributions
of configurational and translational entropy apparently underly the
differences seen between figures \ref{figure07} for the unknot, and
figures \ref{figure14} and \ref{figure15} for the trefoil and figure eight knots,
respectively.  

The analysis of the data for $f_t(\phi)$ proceeded in the same way as for
the unknot.  The minimum free energy was determined by finding the concentration
$\phi^*$ at the minimum of the curves in figures \ref{figure14} and \ref{figure15}, 
and fitting the Flory-Huggins model to data with concentrations $\phi>\phi^*$. 
This gives estimates for the Flory Interaction Parameter $\chi_K$ for compressed knotted
polygons. The results are in the second and third columns of table \ref{tablechift}, and
also plotted in figure \ref{figure06} where the data points \scalebox{0.7}{$\lozenge$} 
correspond to the results for the trefoil knot $3_1$, and the data points
\scalebox{0.7}{$\square$} correspond to the results for the figure eight knot $4_1$.  
These results indicate that the estimate of the Flory interaction parameter $\chi_K$ for
knotted polygons is, in the limit that $L\to\infty$, the same as for the unknot
and that equation \Ref{eqn18A} remains the best estimate in this study.

\begin{figure}[t]
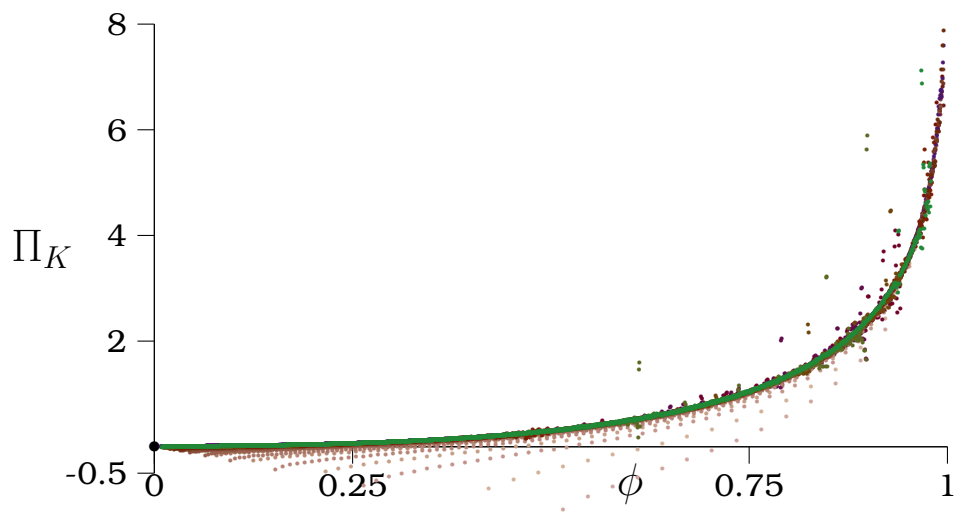

\input Figures/figure17.tex
\caption{Calculated osmotic pressures $\Pi_{K}$ for compressed lattice knots of types
$3_1$ and $4_1$.  On the scale in this picture these data are not distinguishable, and
the osmotic pressure appears to be independent of knot type for high concentrations.
For low concentrations, however, the osmotic pressure is negative, and appears to 
be a function of knot type.}
\label{figure17}  
\end{figure}

\begin{figure}[t]
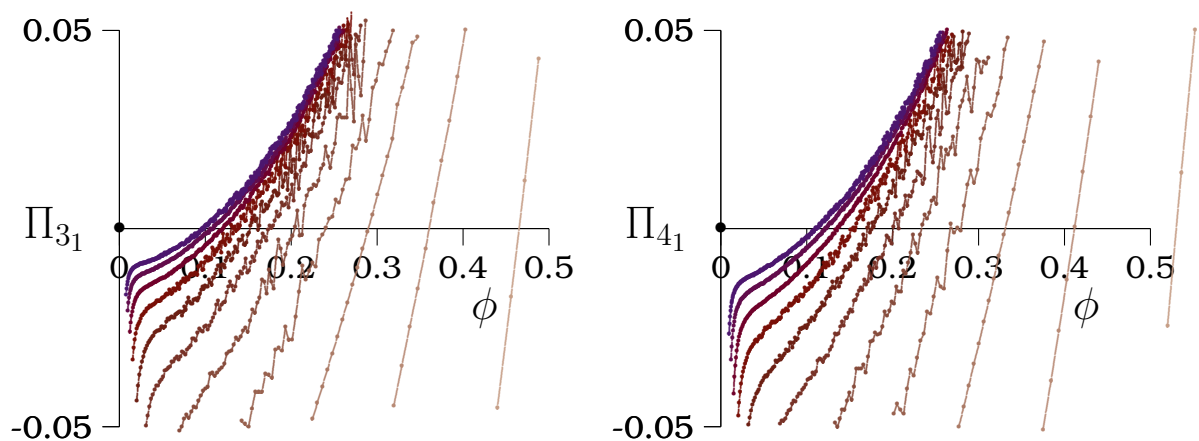

\input Figures/figure19.tex
\caption{The osmotic pressures $\Pi_{3_1}$ (left panel) and 
$\Pi_{4_1}$ (right panel) for compressed 
lattice knots of knot types $3_1$ and $4_1$.  Both pressures are negative at
low concentration, passing through critical concentration $\phi_0$ where the
pressure is zero, to positive osmotic pressure for higher concentration.}
\label{figure19}  
\end{figure}

The osmotic pressure of compressed lattice knots of types $3_1$ and $4_1$
can also be computed using a numerical derivative to approximate equation
\Ref{eqn19B}.  In figure \ref{figure17} the results are shown for both knot
types, and on the scale of the figure there are little differences between these
knots, and with the result in figure \ref{figure09} for the unknot.  However, closer
examination at low concentration shows that the osmotic pressure is negative.
This is shown in figure \ref{figure19}, and these results should be compared to
the result for the unknot in figure \ref{figure10}.

\section{Conclusions}
\label{section4}

The mean field expressions for the free energies in Flory-Huggins theory are
good approximations to the free energy of a model of compressed lattice
polygons in a cube of dimension $L$ at concentrations sufficiently larger than
zero.  This is, for example, seen in the data plotted in figure \ref{figure05} 
and in figure \ref{figure07}.  It was also possible to determine a consistent 
value of the Flory Interaction Parameter $\chi$ from the data, as shown in 
figure \ref{figure06}.  The estimate $\chi_K \approx 0.18$ for $K\in\{0_1,3_1,4_1\}$
appears to be independent of knot type, and also shows, in the contect of
Flory-Huggins theory, that the cubic lattice acts as a ``good solvent" for lattice knots
(since $\chi_K>0$ the solvent is not athermal, and since $\chi_K<\sfrac{1}{2}$ it is
also not a $\theta$-solvent).

There are, however, some additional observations which can be made about
using Flory-Huggins theory to model the free energy of compressed lattice knots.
These are:
\begin{itemize}
\item The numerical results support a conjecture that the Flory Interaction Parameter is
independent of knot type;
\item While the Flory-Huggins expressions in equation \Ref{eqn15} and
equation \Ref{eqn17} are good approximations of the free energies
$F_V$ and $f_t$, the Flory-Huggins expression for the osmotic pressure 
deviates from the calculated data in figures \ref{figure09} and \ref{figure17}, 
in particular at concentrations close to $1$;
\item On magnification of the osmotic pressure in figures \ref{figure10} and
\ref{figure19},  Flory-Huggins theory is not a good approximation at low 
concentrations -- it fails to explain the dependence of osmotic pressure on 
knot type, and does not account for the measured negative values of the osmotic pressure 
in the data.
\end{itemize}

\vspace{1cm}
\noindent{\bf Acknowledgements:} EJJvR acknowledges financial support 
from NSERC (Canada) in the form of Discovery Grant RGPIN-2014-04731.  
I am grateful to SG Whittington for helpful remarks.

\vspace{2cm}
\noindent{\bf References}
\bibliographystyle{plain}
\bibliography{References}

\end{document}

%% file: macros.tex
\def\2;{\;\;}

\def\IntZ{{\mathbb Z}}

\def\Ref#1{(\ref{#1})}

\def\Sfrac#1#2{\hbox{\large $\frac{#1}{#2}$}}
\def\sfrac#1#2{\hbox{\nor $\frac{#1}{#2}$}}

\def\nor{\normalsize}
\def\fns{\scriptsize}

\def\tiny#1{\scalebox{0.75}{#1}}
\def\Tiny#1{\scalebox{0.5}{#1}}

%% file: macros-pictex.tex
\definecolor{blue}{rgb}{0,0.18,0.39}
\definecolor{RoyalBlue}{rgb}{0,0.2,0.7}

\def\axes#1#2#3#4#5#6#7{
\setplotarea x from #7 to #5, y from #2 to #6
\setplotarea x from #1 to #5, y from #2 to #6
\axis left shiftedto x=#3 
        ticks
        withvalues #6 #2 /
        at  #6 #2 /
 /
\axis bottom shiftedto y=#4
        ticks
        withvalues #1 #5 /
        at  #1 #5 /
/
\put {\footnotesize$\bullet$} at #3 #4
}

\def\axesnolabels#1#2#3#4#5#6#7{
\setplotarea x from #7 to #5, y from #2 to #6
\setplotarea x from #1 to #5, y from #2 to #6
\axis left shiftedto x=#3 
        ticks
        at  #6 #2 /
 /
\axis bottom shiftedto y=#4
        ticks
        at  #1 #5 /
/
\put {\footnotesize$\bullet$} at #3 #4
}

\definecolor{Maroon}{cmyk}{0,0.87,0.68,0.62}
\definecolor{Brown}{rgb}{0.7,0.3,0}
\definecolor{Navy}{rgb}{0.3,0.0,0.4}
\definecolor{Red}{cmyk}{0,1,1,0}
\definecolor{BrickRed}{cmyk}{0.16,0.89,0.61,0.02}
\definecolor{DarkRed}{cmyk}{0,1,1,0.5}
\definecolor{DarkBlue}{cmyk}{1,1,0,0.2}
\definecolor{DarkGreen}{cmyk}{1,0,1,0.4}
\definecolor{Green}{cmyk}{1,0,1,0}
\definecolor{DarkBrown}{cmyk}{0,0.81,1,0.6}
\definecolor{OrangeRed}{cmyk}{0,1,0.87,0}
\definecolor{RedOrange}{cmyk}{0,0.77,0.87,0}
\definecolor{Orange}{cmyk}{0,0.61,0.87,0}
\definecolor{Offwhite}{rgb}{.8,0.9,.8}
\definecolor{Offwhite2}{cmyk}{.04,.02,.01,0}
\definecolor{Tan}{rgb}{0.82,0.70,0.55}
\definecolor{Blue}{rgb}{0,0,1}
\definecolor{RoyalBlue}{rgb}{0.25,0.41,0.88}
\definecolor{Sepia}{rgb}{0.37,0.14,0.07}
\definecolor{myblue}{cmyk}{0.025,0.05,0,0}
\definecolor{Mahogany}{cmyk}{0.18,0.87,1,0.08}

\definecolor{green1}{cmyk}{0.25,0,0.76,0}
\definecolor{green2}{cmyk}{0.25,0,0.76,0.07}
\definecolor{green3}{cmyk}{0.25,0,0.76,0.20}
\definecolor{green4}{cmyk}{0.25,0,0.75,0.30}
\definecolor{green5}{cmyk}{0.25,0,0.75,0.40}
\definecolor{green6}{cmyk}{0.25,0,0.75,0.50}

\definecolor{B02}{cmyk}{0,0.14,0.22,0.12}
\definecolor{B03}{cmyk}{0,0.16,0.26,0.16}
\definecolor{B04}{cmyk}{0,0.19,0.28,0.19}
\definecolor{B05}{cmyk}{0,0.25,0.32,0.25}
\definecolor{B06}{cmyk}{0,0.31,0.36,0.31}
\definecolor{B07}{cmyk}{0,0.37,0.40,0.37}
\definecolor{B08}{cmyk}{0,0.46,0.46,0.46}
\definecolor{B09}{cmyk}{0,0.55,0.52,0.54}
\definecolor{B10}{cmyk}{0,0.69,0.61,0.62}
\definecolor{B11}{cmyk}{0,0.78,0.70,0.68}
\definecolor{B12}{cmyk}{0,0.93,0.85,0.60}
\definecolor{B13}{cmyk}{0.25,1,0.6,0.50}
\definecolor{B14}{cmyk}{0.5,1,0.30,0.40}
\definecolor{B15}{cmyk}{0.75,1,0,0.30}

\definecolor{C02}{cmyk}{0,0.22,0.14,0.12}
\definecolor{C03}{cmyk}{0,0.26,0.16,0.16}
\definecolor{C04}{cmyk}{0,0.28,0.19,0.19}
\definecolor{C05}{cmyk}{0,0.32,0.25,0.25}
\definecolor{C06}{cmyk}{0,0.36,0.31,0.31}
\definecolor{C07}{cmyk}{0,0.40,0.37,0.37}
\definecolor{C08}{cmyk}{0,0.46,0.46,0.46}
\definecolor{C09}{cmyk}{0,0.52,0.55,0.54}
\definecolor{C10}{cmyk}{0,0.61,0.69,0.62}
\definecolor{C11}{cmyk}{0,0.70,0.78,0.68}
\definecolor{C12}{cmyk}{0,0.85,0.93,0.60}
\definecolor{C13}{cmyk}{0.25,0.60,1,0.50}
\definecolor{C14}{cmyk}{0.5,0.30,1,0.40}
\definecolor{C15}{cmyk}{0.75,0,1,0.30}

\def\colM{\color{Maroon}}

%% file: Figures/figure01.tex
\beginpicture
\setcoordinatesystem units <1.75pt,1.75pt>
\setplotarea x from -40 to 60, y from  0 to 100

\put { } at -120 0

\color{red}
\put {\beginpicture \grid 10 10  \endpicture} at 0 0

\setcoordinatesystem units <1.75pt,1.75pt> point at 40 0 
\setplotarea x from 0 to 75, y from  0 to 100
\color{red}
\setplotsymbol ({\Large$\bullet$})
\plot -3 -3 103 -3 103 103 -3 103 -3 -3 /

\setplotsymbol ({\large$\bullet$})
\color{RoyalBlue}
\plot 40 40 30 40 30 50 20 50 10 50 0 50 0 40 0 30 0 20 0 10 
0 0 10 0 20 0 20 10 10 10 10 20 10 30 10 40 20 40 20 30 30 30 40 30  
40 20 30 20 30 10 40 10 50 10 60 10 60 0 70 0 70 10 80 10 80 20 90 20 90 30
90 40 100 40 100 50 100 60 90 60 80 60 70 60 70 50 60 50 60 60 60 70 50 70 
40 70 40 80 40 90 40 100 30 100 30 90 30 80 20 80 20 70 20 60 30 60 40 60 
40 50 50 50  50 40 40 40  /

\color{blue}
\multiput {{\LARGE$\bullet$}} at 
40 40 30 40 30 50 20 50 10 50 0 50 0 40 0 30 0 20 0 10 
0 0 10 0 20 0 20 10 10 10 10 20 10 30 10 40 20 40 20 30 30 30 40 30  
40 20 30 20 30 10 40 10 50 10 60 10 60 0 70 0 70 10 80 10 80 20 90 20 90 30
90 40 100 40 100 50 100 60 90 60 80 60 70 60 70 50 60 50 60 60 60 70 50 70 
40 70 40 80 40 90 40 100 30 100 30 90 30 80 20 80 20 70 20 60 30 60 40 60 
40 50 50 50  50 40  /

\color{black}
\normalcolor

\endpicture

%% file: Figures/figure02-1.tex
\beginpicture
\setplotarea x from -75 to 75, y from -100 to 100

\setplotsymbol ({$\bullet$})

\color{red}
\plot -75 -75 75 -75 75 75 -75 75 -75 -75 /

\color{blue}
\circulararc 360 degrees from -55 -25 center at -50 -25 

\setplotsymbol ({\Tiny{$\bullet$}})
\color{black}
\arrow <10pt>  [.2,.67] from -40 -15 to -20 5

\normalcolor

\put {Translational Entropy} at 0 -90

\endpicture

%% file: Figures/figure02-2.tex
\beginpicture
\setplotarea x from -75 to 75, y from -100 to 100

\setplotsymbol ({$\bullet$})

\color{red}
\plot -75 -75 75 -75 75 75 -75 75 -75 -75 /

\color{blue}
\plot -65 -60 -65 -5 /
\circulararc 90 degrees from -60 0 center at -60 -5 
\plot -60 0 10 0 /
\circulararc 180 degrees from 10 0 center at 10 5
\plot -60 10 10 10 /
\circulararc 180 degrees from -60 20 center at -60 15
\plot -60 20 10 20 /
\circulararc 180 degrees from 10 20 center at 10 25
\plot 10 30 -40 30 /
\circulararc 180 degrees from -40 40 center at -40 35
\circulararc 90 degrees from -40 40 center at -40 45
\plot -35 45 -35 60 /
\circulararc 90 degrees from -30 65 center at -30 60
\plot -30 65 45 65 /
\circulararc 90 degrees from 50 60 center at 45 60
\plot 50 60 50 20 /
\circulararc 90 degrees from 50 20 center at 55 20
\plot 60 15 55 15 /
\circulararc 90 degrees from 65 10 center at 60 10
\plot 65 10 65 -20 /
\circulararc 90 degrees from 60 -25 center at 60 -20
\plot 60 -25 20 -25 /
\circulararc 180 degrees from 20 -25 center at 20 -30 
\plot 20 -35 30 -35 /
\circulararc 180 degrees from 30 -45 center at 30 -40 
\circulararc 180 degrees from 30 -45 center at 30 -50 
\circulararc 180 degrees from 30 -65 center at 30 -60 
\plot 30 -65 -40 -65 /
\circulararc 90 degrees from -45 -60 center at -40 -60 
\circulararc 180 degrees from -45 -60 center at -50 -60 
\circulararc 180 degrees from -65 -60 center at -60 -60 

\setplotsymbol ({\Tiny{$\bullet$}})
\color{black}

\normalcolor

\put {Conformational Entropy} at 0 -90

\endpicture

%% file: Figures/figure02-3.tex
\beginpicture
\setplotarea x from -75 to 75, y from -100 to 100

\setplotsymbol ({$\bullet$})

\color{red}
\plot -75 -75 75 -75 75 75 -75 75 -75 -75 /

\color{blue}
\setquadratic
\plot 20 60 0 50 -15 20 -15 -15 5 -20 25 -10 35 3 43 25 45 42   /
\plot 45 58  35 62  22 52  20 40  35 32  /
\plot 52 30  58 30  63 33  65 45  45 50  40 50  35 52 /

\setlinear

\setplotsymbol ({\Tiny{$\bullet$}})
\color{black}
\arrow <10pt>  [.2,.67] from 60 20 to 55 -10

\normalcolor

\put {Topological Entropy} at 0 -90

\endpicture

%% file: Figures/figure03.tex
\beginpicture
\setcoordinatesystem units <1pt,1pt>
\setplotarea x from -20 to 420, y from -10 to 110
\setplotarea x from 0 to 400, y from 0 to 100

\setplotsymbol ({\footnotesize{$\bullet$}})

\setquadratic
\color{red}
\ellipticalarc axes ratio 8:10  360 degrees from 40 0 center at 40 50 
\color{black}
\normalcolor

\put {\Large$0_1$} at 50 -12 

\color{red}
\setcoordinatesystem units <1pt,1pt> point at -135 0
\plot 92 52 80 35 70 24 60 15 50 8 40 2 30 0 /
\plot 30 0 20 2 10 12 0 30 0 50 10 70 30 80  / 
\plot 50 80 60 78 70 75 80 70 90 65 102 55  110 40 113 20 108 5 95 0 80 5 75 10 70 15 /
\plot 60 27 50 40 40 70 40 80 42 90 45 100 50 105 65 110 80 107 90 103 100 93 104 80 100 70   /
\color{black}
\normalcolor

\put {\Large$3_1$} at 65 -10 

\color{red}
\setcoordinatesystem units <1pt,1pt> point at -300 0
\plot 55 50 65 40 67 30 65 20 60 12 50 4 40 0  30 2  20 7 10 15 0 30 -5 52 0 68 7 75 20 80 /
\plot 35 80 40 81 50 81.5 60 81 70 79 80 75 90 65 100 45 100 25 97 18  93 10 83 0 65 4  /
\plot 50 15 45 20 42 26 40 37 45 50 50 58 58 64 63 68 65 72  /
\plot 66 88  65 97 60 104 55 108 50 108  40 105 30 95  28 85 31 70  35 65 40 60  /
\color{black}
\normalcolor

\put {\Large$4_1$} at 55 -10 

\endpicture

%% file: Figures/figure04.tex
\beginpicture
\color{black}
\normalcolor
\color{black}

\setplotsymbol ({\LARGE$\cdot$})

\setcoordinatesystem units <225pt,285pt>
\setplotarea x from -0.4 to 1.1, y from -0.45 to 0.1
\setplotarea x from 0 to 1.1, y from -0.45 to 0.1
\axesnolabels{0}{-0.45}{0}{0}{1}{0.1}{-0.0}
\axis left shiftedto x=0 label {\Large\stack{$F_V$}} /
\axis bottom shiftedto y=0 label {\hspace{4cm}\Large$\phi$} /

\plot 0 0 1.1 0 /
\plot 0 -0.45 0 0.1 /

\colM
\setplotsymbol ({\fns$\bullet$})
\color{DarkGreen}
\plot 0 0  0.025 -0.0303  0.050 -0.0568  0.075 -0.0817  0.100 -0.1054  0.125 -0.1280  0.150 -0.1497  0.175 -0.1706  0.200 -0.1907  0.225 -0.2100  0.250 -0.2285  0.275 -0.2463  0.300 -0.2633  0.325 -0.2795  0.350 -0.2949  0.375 -0.3094  0.400 -0.3231  0.425 -0.3360  0.450 -0.3479  0.475 -0.3588  0.500 -0.3687  0.525 -0.3776  0.550 -0.3853  0.575 -0.3919  0.600 -0.3972  0.625 -0.4011  0.650 -0.4036  0.675 -0.4045  0.700 -0.4037  0.725 -0.4010  0.750 -0.3963  0.775 -0.3893  0.800 -0.3797  0.825 -0.3673  0.850 -0.3515  0.875 -0.3318  0.900 -0.3072  0.925 -0.2766  0.950 -0.2378  0.975 -0.1861  1.000 -0.1000   /


\color{black}
\normalcolor
\color{black}

\endpicture

%% file: Figures/figure06.tex
\beginpicture

\color{black}
\setcoordinatesystem units <700pt,125pt>
\axes{0}{0}{0}{0}{0.4}{1.3}{0}
\axis left shiftedto x=0 label {\Large$\chi_K$} 
      ticks
      withvalues 0.25 0.5 0.75 1.0 /
      at 0.25 0.5 0.75 1.0  /
 /  
\axis bottom shiftedto y=0 
    ticks
    withvalues 0.2 /
    at 0.2  /
/

\put {\Large$\frac{1}{L}$} at 0.3 -0.12


\color{red}
\multiput {$\bullet$} at 0.333 0.4372  0.25 0.3611  0.20 0.3084  0.1667 0.2810  0.1429 0.2617
0.125 0.2514  0.1111 0.2437  0.10 0.2388  0.0909 0.2335  0.0833 0.2284 0.0769 0.2188 0.0714  0.2122 0.0667 0.2176 /

\multiput{$\circ$} at 0.333 0.4265  0.25 0.3530  0.20 0.3046  0.1667 0.2748  0.1429 0.2561
0.125 0.2437  0.1111 0.2370  0.10 0.2320  0.0909 0.2260  0.0833 0.2220 0.0769 0.2165 0.0714  0.2192 0.0667 0.2132 /

\color{blue}
\multiput {\scalebox{0.7}{$\blacklozenge$}} at 0.25 0.9796  0.20 0.5983  0.1667 0.4185  0.1429 0.3333
0.125 0.2905  0.1111 0.2662  0.10 0.2524  0.0909 0.2433  0.0833 0.2367 0.0769 0.2235 0.0714  0.2165 0.0667 0.2246 /

\multiput{\scalebox{0.7}{$\lozenge$}} at  0.25 0.6906  0.20 0.4482  0.1667 0.3564  0.1429 0.3053
0.125 0.2760  0.1111 0.2569  0.10 0.2471  0.0909 0.2383  0.0833 0.2325 0.0769 0.2231 0.0714  0.2180 0.0667 0.2193 /

\color{Maroon}
\multiput {\scalebox{0.5}{$\blacksquare$}} at 0.25 1.2820  0.20 0.7395  0.1667 0.4913  0.1429 0.3737
0.125 0.3189  0.1111 0.2814  0.10 0.2619  0.0909 0.2501  0.0833 0.2396 0.0769 0.2272 0.0714  0.2203 0.0667 0.2239 /

\multiput{\scalebox{0.5}{$\square$}} at  0.25 0.8784  0.20 0.5115  0.1667 0.3886  0.1429 0.3265
0.125 0.2895  0.1111 0.2686  0.10 0.2531  0.0909 0.2450  0.0833 0.2370 0.0769 0.2270 0.0714  0.2225 0.0667 0.2218 /

\color{black}
\put{\underline{$f_t$}:} at 0.38 1.01
\color{Maroon}
\put {$4_1$ --  \scalebox{0.5}{$\square$}} at 0.4 0.9
\color{blue}
\put {$3_1$ --  \scalebox{0.7}{$\lozenge$}} at 0.4 0.8
\color{red}
\put {$0_1$ --  $\circ$} at 0.4 0.7

\color{black}
\put{\underline{$F_V$}:} at 0.38 0.45
\color{Maroon}
\put {$4_1$ --  \scalebox{0.5}{$\blacksquare$}} at 0.4 0.34
\color{blue}
\put {$3_1$ --  \scalebox{0.7}{$\blacklozenge$}} at 0.4 0.24
\color{red}
\put {$0_1$ --  $\bullet$} at 0.4 0.14

\color{black}
\normalcolor

\endpicture

%% file: Figures/figure08.tex
\beginpicture

\normalcolor
\color{black}
\setcoordinatesystem units <275pt,120pt>
\axes{0}{-1.7}{0}{0}{1}{0}{0}
\axis left shiftedto x=0 label {\Large$f_t(\phi)$} 
      ticks
      withvalues -0.5 -1.0 /
      at -0.5 -1.0 /
 /  
\axis bottom shiftedto y=0 
    ticks
    withvalues 0.25 0.75 /
    at 0.25 0.75  /
/

\put {\Large$\phi$} at 0.6 -0.12

\color{B09}
\plot 
0.0082 -1.5533  0.0110 -1.4285  0.0137 -1.3760  0.0165 -1.3506  0.0192 -1.3389  0.0219 -1.3338  0.0247 -1.3318 
0.0274 -1.3322  0.0302 -1.3336  0.0329 -1.3353  0.0357 -1.3374  0.0384 -1.3403  0.0412 -1.3431  0.0439 -1.3459  0.0466 -1.3482  0.0494 -1.3504  0.0521 -1.3527  0.0549 -1.3547  0.0576 -1.3565  0.0604 -1.3580  0.0631 -1.3596  0.0658 -1.3609  0.0686 -1.3627  0.0713 -1.3638  0.0741 -1.3649  0.0768 -1.3658  0.0796 -1.3666  0.0823 -1.3673  0.0850 -1.3677  0.0878 -1.3681  0.0905 -1.3683  0.0933 -1.3687  0.0960 -1.3690  0.0988 -1.3692  0.1015 -1.3694  0.1043 -1.3697  0.1070 -1.3694  0.1097 -1.3697  0.1125 -1.3693  0.1152 -1.3692  0.1180 -1.3691  0.1207 -1.3687  0.1235 -1.3686  0.1262 -1.3683  0.1289 -1.3679  0.1317 -1.3675  0.1344 -1.3671  0.1372 -1.3667  0.1399 -1.3662  0.1427 -1.3657  0.1454 -1.3652  0.1481 -1.3647  0.1509 -1.3643  0.1536 -1.3639  0.1564 -1.3634  0.1591 -1.3630  0.1619 -1.3623  0.1646 -1.3616  0.1674 -1.3609  0.1701 -1.3601  0.1728 -1.3594  0.1756 -1.3586  0.1783 -1.3579  0.1811 -1.3571  0.1838 -1.3564  0.1866 -1.3556  0.1893 -1.3549  0.1920 -1.3543  0.1948 -1.3533  0.1975 -1.3525  0.2003 -1.3517  0.2030 -1.3509  0.2058 -1.3501  0.2085 -1.3492  0.2112 -1.3484  0.2140 -1.3477  0.2167 -1.3467  0.2195 -1.3458  0.2222 -1.3451  0.2250 -1.3441  0.2277 -1.3433  0.2305 -1.3427  0.2332 -1.3417  0.2359 -1.3408  0.2387 -1.3399  0.2414 -1.3391  0.2442 -1.3382  0.2469 -1.3371  0.2497 -1.3361  0.2524 -1.3351  0.2551 -1.3342  0.2579 -1.3332  0.2606 -1.3323  0.2634 -1.3312  0.2661 -1.3302  0.2689 -1.3292  0.2716 -1.3282  0.2743 -1.3273  0.2771 -1.3263  0.2798 -1.3253  0.2826 -1.3243  0.2853 -1.3233  0.2881 -1.3221  0.2908 -1.3211  0.2936 -1.3199  0.2963 -1.3189  0.2990 -1.3178  0.3018 -1.3167  0.3045 -1.3155  0.3073 -1.3144  0.3100 -1.3133  0.3128 -1.3123  0.3155 -1.3111  0.3182 -1.3099  0.3210 -1.3087  0.3237 -1.3078  0.3265 -1.3067  0.3292 -1.3056  0.3320 -1.3045  0.3347 -1.3035  0.3374 -1.3023  0.3402 -1.3012  0.3429 -1.3000  0.3457 -1.2988  0.3484 -1.2976  0.3512 -1.2964  0.3539 -1.2952  0.3567 -1.2940  0.3594 -1.2927  0.3621 -1.2915  0.3649 -1.2904  0.3676 -1.2891  0.3704 -1.2880  0.3731 -1.2867  0.3759 -1.2856  0.3786 -1.2843  0.3813 -1.2831  0.3841 -1.2819  0.3868 -1.2806  0.3896 -1.2793  0.3923 -1.2780  0.3951 -1.2767  0.3978 -1.2754  0.4005 -1.2741  0.4033 -1.2728  0.4060 -1.2715  0.4088 -1.2702  0.4115 -1.2688  0.4143 -1.2675  0.4170 -1.2662  0.4198 -1.2648  0.4225 -1.2635  0.4252 -1.2621  0.4280 -1.2606  0.4307 -1.2593  0.4335 -1.2579  0.4362 -1.2565  0.4390 -1.2551  0.4417 -1.2537  0.4444 -1.2523  0.4472 -1.2509  0.4499 -1.2495  0.4527 -1.2481  0.4554 -1.2466  0.4582 -1.2452  0.4609 -1.2437  0.4636 -1.2423  0.4664 -1.2408  0.4691 -1.2393  0.4719 -1.2379  0.4746 -1.2364  0.4774 -1.2349  0.4801 -1.2334  0.4829 -1.2319  0.4856 -1.2304  0.4883 -1.2289  0.4911 -1.2274  0.4938 -1.2259  0.4966 -1.2244  0.4993 -1.2228  0.5021 -1.2213  0.5048 -1.2198  0.5075 -1.2182  0.5103 -1.2166  0.5130 -1.2150  0.5158 -1.2135  0.5185 -1.2119  0.5213 -1.2103  0.5240 -1.2087  0.5267 -1.2072  0.5295 -1.2055  0.5322 -1.2039  0.5350 -1.2023  0.5377 -1.2007  0.5405 -1.1991  0.5432 -1.1974  0.5460 -1.1958  0.5487 -1.1941  0.5514 -1.1924  0.5542 -1.1908  0.5569 -1.1891  0.5597 -1.1874  0.5624 -1.1857  0.5652 -1.1840  0.5679 -1.1823  0.5706 -1.1806  0.5734 -1.1788  0.5761 -1.1771  0.5789 -1.1753  0.5816 -1.1736  0.5844 -1.1718  0.5871 -1.1701  0.5898 -1.1683  0.5926 -1.1665  0.5953 -1.1647  0.5981 -1.1629  0.6008 -1.1611  0.6036 -1.1593  0.6063 -1.1574  0.6091 -1.1556  0.6118 -1.1538  0.6145 -1.1519  0.6173 -1.1500  0.6200 -1.1482  0.6228 -1.1463  0.6255 -1.1443  0.6283 -1.1424  0.6310 -1.1405  0.6337 -1.1386  0.6365 -1.1367  0.6392 -1.1348  0.6420 -1.1328  0.6447 -1.1309  0.6475 -1.1289  0.6502 -1.1269  0.6529 -1.1250  0.6557 -1.1230  0.6584 -1.1210  0.6612 -1.1189  0.6639 -1.1169  0.6667 -1.1149  0.6694 -1.1128  0.6722 -1.1108  0.6749 -1.1087  0.6776 -1.1066  0.6804 -1.1045  0.6831 -1.1024  0.6859 -1.1003  0.6886 -1.0982  0.6914 -1.0960  0.6941 -1.0939  0.6968 -1.0917  0.6996 -1.0895  0.7023 -1.0873  0.7051 -1.0851  0.7078 -1.0829  0.7106 -1.0807  0.7133 -1.0784  0.7160 -1.0762  0.7188 -1.0739  0.7215 -1.0716  0.7243 -1.0693  0.7270 -1.0670  0.7298 -1.0647  0.7325 -1.0623  0.7353 -1.0600  0.7380 -1.0576  0.7407 -1.0552  0.7435 -1.0528  0.7462 -1.0504  0.7490 -1.0479  0.7517 -1.0455  0.7545 -1.0430  0.7572 -1.0405  0.7599 -1.0380  0.7627 -1.0355  0.7654 -1.0330  0.7682 -1.0304  0.7709 -1.0279  0.7737 -1.0253  0.7764 -1.0227  0.7791 -1.0200  0.7819 -1.0174  0.7846 -1.0147  0.7874 -1.0120  0.7901 -1.0093  0.7929 -1.0066  0.7956 -1.0039  0.7984 -1.0011  0.8011 -0.9983  0.8038 -0.9955  0.8066 -0.9927  0.8093 -0.9898  0.8121 -0.9869  0.8148 -0.9840  0.8176 -0.9812  0.8203 -0.9782  0.8230 -0.9753  0.8258 -0.9723  0.8285 -0.9693  0.8313 -0.9663  0.8340 -0.9632  0.8368 -0.9601  0.8395 -0.9570  0.8422 -0.9538  0.8450 -0.9507  0.8477 -0.9474  0.8505 -0.9442  0.8532 -0.9410  0.8560 -0.9377  0.8587 -0.9344  0.8615 -0.9310  0.8642 -0.9277  0.8669 -0.9242  0.8697 -0.9208  0.8724 -0.9173  0.8752 -0.9138  0.8779 -0.9102  0.8807 -0.9066  0.8834 -0.9030  0.8861 -0.8993  0.8889 -0.8956  0.8916 -0.8919  0.8944 -0.8881  0.8971 -0.8842  0.8999 -0.8803  0.9026 -0.8764  0.9053 -0.8723  0.9081 -0.8683  0.9108 -0.8642  0.9136 -0.8600  0.9163 -0.8557  0.9191 -0.8513  0.9218 -0.8469  0.9246 -0.8425  0.9273 -0.8380  0.9300 -0.8335  0.9328 -0.8288  0.9355 -0.8241  0.9383 -0.8193  0.9410 -0.8144  0.9438 -0.8094  0.9465 -0.8043  0.9492 -0.7991  0.9520 -0.7939  0.9547 -0.7885  0.9575 -0.7831  0.9602 -0.7774  0.9630 -0.7713  0.9657 -0.7653  0.9684 -0.7593  0.9712 -0.7530  0.9739 -0.7465  0.9767 -0.7398  0.9794 -0.7329  0.9822 -0.7257  0.9849 -0.7182  0.9877 -0.7103  0.9904 -0.7022  0.9931 -0.6935  0.9959 -0.6840  0.9986 -0.6744  /
\multiput {\large$\cdot$} at 
0.0082 -1.5533  0.0110 -1.4285  0.0137 -1.3760  0.0165 -1.3506  0.0192 -1.3389  0.0219 -1.3338  0.0247 -1.3318 
0.0274 -1.3322  0.0302 -1.3336  0.0329 -1.3353  0.0357 -1.3374  0.0384 -1.3403  0.0412 -1.3431  0.0439 -1.3459  0.0466 -1.3482  0.0494 -1.3504  0.0521 -1.3527  0.0549 -1.3547  0.0576 -1.3565  0.0604 -1.3580  0.0631 -1.3596  0.0658 -1.3609  0.0686 -1.3627  0.0713 -1.3638  0.0741 -1.3649  0.0768 -1.3658  0.0796 -1.3666  0.0823 -1.3673  0.0850 -1.3677  0.0878 -1.3681  0.0905 -1.3683  0.0933 -1.3687  0.0960 -1.3690  0.0988 -1.3692  0.1015 -1.3694  0.1043 -1.3697  0.1070 -1.3694  0.1097 -1.3697  0.1125 -1.3693  0.1152 -1.3692  0.1180 -1.3691  0.1207 -1.3687  0.1235 -1.3686  0.1262 -1.3683  0.1289 -1.3679  0.1317 -1.3675  0.1344 -1.3671  0.1372 -1.3667  0.1399 -1.3662  0.1427 -1.3657  0.1454 -1.3652  0.1481 -1.3647  0.1509 -1.3643  0.1536 -1.3639  0.1564 -1.3634  0.1591 -1.3630  0.1619 -1.3623  0.1646 -1.3616  0.1674 -1.3609  0.1701 -1.3601  0.1728 -1.3594  0.1756 -1.3586  0.1783 -1.3579  0.1811 -1.3571  0.1838 -1.3564  0.1866 -1.3556  0.1893 -1.3549  0.1920 -1.3543  0.1948 -1.3533  0.1975 -1.3525  0.2003 -1.3517  0.2030 -1.3509  0.2058 -1.3501  0.2085 -1.3492  0.2112 -1.3484  0.2140 -1.3477  0.2167 -1.3467  0.2195 -1.3458  0.2222 -1.3451  0.2250 -1.3441  0.2277 -1.3433  0.2305 -1.3427  0.2332 -1.3417  0.2359 -1.3408  0.2387 -1.3399  0.2414 -1.3391  0.2442 -1.3382  0.2469 -1.3371  0.2497 -1.3361  0.2524 -1.3351  0.2551 -1.3342  0.2579 -1.3332  0.2606 -1.3323  0.2634 -1.3312  0.2661 -1.3302  0.2689 -1.3292  0.2716 -1.3282  0.2743 -1.3273  0.2771 -1.3263  0.2798 -1.3253  0.2826 -1.3243  0.2853 -1.3233  0.2881 -1.3221  0.2908 -1.3211  0.2936 -1.3199  0.2963 -1.3189  0.2990 -1.3178  0.3018 -1.3167  0.3045 -1.3155  0.3073 -1.3144  0.3100 -1.3133  0.3128 -1.3123  0.3155 -1.3111  0.3182 -1.3099  0.3210 -1.3087  0.3237 -1.3078  0.3265 -1.3067  0.3292 -1.3056  0.3320 -1.3045  0.3347 -1.3035  0.3374 -1.3023  0.3402 -1.3012  0.3429 -1.3000  0.3457 -1.2988  0.3484 -1.2976  0.3512 -1.2964  0.3539 -1.2952  0.3567 -1.2940  0.3594 -1.2927  0.3621 -1.2915  0.3649 -1.2904  0.3676 -1.2891  0.3704 -1.2880  0.3731 -1.2867  0.3759 -1.2856  0.3786 -1.2843  0.3813 -1.2831  0.3841 -1.2819  0.3868 -1.2806  0.3896 -1.2793  0.3923 -1.2780  0.3951 -1.2767  0.3978 -1.2754  0.4005 -1.2741  0.4033 -1.2728  0.4060 -1.2715  0.4088 -1.2702  0.4115 -1.2688  0.4143 -1.2675  0.4170 -1.2662  0.4198 -1.2648  0.4225 -1.2635  0.4252 -1.2621  0.4280 -1.2606  0.4307 -1.2593  0.4335 -1.2579  0.4362 -1.2565  0.4390 -1.2551  0.4417 -1.2537  0.4444 -1.2523  0.4472 -1.2509  0.4499 -1.2495  0.4527 -1.2481  0.4554 -1.2466  0.4582 -1.2452  0.4609 -1.2437  0.4636 -1.2423  0.4664 -1.2408  0.4691 -1.2393  0.4719 -1.2379  0.4746 -1.2364  0.4774 -1.2349  0.4801 -1.2334  0.4829 -1.2319  0.4856 -1.2304  0.4883 -1.2289  0.4911 -1.2274  0.4938 -1.2259  0.4966 -1.2244  0.4993 -1.2228  0.5021 -1.2213  0.5048 -1.2198  0.5075 -1.2182  0.5103 -1.2166  0.5130 -1.2150  0.5158 -1.2135  0.5185 -1.2119  0.5213 -1.2103  0.5240 -1.2087  0.5267 -1.2072  0.5295 -1.2055  0.5322 -1.2039  0.5350 -1.2023  0.5377 -1.2007  0.5405 -1.1991  0.5432 -1.1974  0.5460 -1.1958  0.5487 -1.1941  0.5514 -1.1924  0.5542 -1.1908  0.5569 -1.1891  0.5597 -1.1874  0.5624 -1.1857  0.5652 -1.1840  0.5679 -1.1823  0.5706 -1.1806  0.5734 -1.1788  0.5761 -1.1771  0.5789 -1.1753  0.5816 -1.1736  0.5844 -1.1718  0.5871 -1.1701  0.5898 -1.1683  0.5926 -1.1665  0.5953 -1.1647  0.5981 -1.1629  0.6008 -1.1611  0.6036 -1.1593  0.6063 -1.1574  0.6091 -1.1556  0.6118 -1.1538  0.6145 -1.1519  0.6173 -1.1500  0.6200 -1.1482  0.6228 -1.1463  0.6255 -1.1443  0.6283 -1.1424  0.6310 -1.1405  0.6337 -1.1386  0.6365 -1.1367  0.6392 -1.1348  0.6420 -1.1328  0.6447 -1.1309  0.6475 -1.1289  0.6502 -1.1269  0.6529 -1.1250  0.6557 -1.1230  0.6584 -1.1210  0.6612 -1.1189  0.6639 -1.1169  0.6667 -1.1149  0.6694 -1.1128  0.6722 -1.1108  0.6749 -1.1087  0.6776 -1.1066  0.6804 -1.1045  0.6831 -1.1024  0.6859 -1.1003  0.6886 -1.0982  0.6914 -1.0960  0.6941 -1.0939  0.6968 -1.0917  0.6996 -1.0895  0.7023 -1.0873  0.7051 -1.0851  0.7078 -1.0829  0.7106 -1.0807  0.7133 -1.0784  0.7160 -1.0762  0.7188 -1.0739  0.7215 -1.0716  0.7243 -1.0693  0.7270 -1.0670  0.7298 -1.0647  0.7325 -1.0623  0.7353 -1.0600  0.7380 -1.0576  0.7407 -1.0552  0.7435 -1.0528  0.7462 -1.0504  0.7490 -1.0479  0.7517 -1.0455  0.7545 -1.0430  0.7572 -1.0405  0.7599 -1.0380  0.7627 -1.0355  0.7654 -1.0330  0.7682 -1.0304  0.7709 -1.0279  0.7737 -1.0253  0.7764 -1.0227  0.7791 -1.0200  0.7819 -1.0174  0.7846 -1.0147  0.7874 -1.0120  0.7901 -1.0093  0.7929 -1.0066  0.7956 -1.0039  0.7984 -1.0011  0.8011 -0.9983  0.8038 -0.9955  0.8066 -0.9927  0.8093 -0.9898  0.8121 -0.9869  0.8148 -0.9840  0.8176 -0.9812  0.8203 -0.9782  0.8230 -0.9753  0.8258 -0.9723  0.8285 -0.9693  0.8313 -0.9663  0.8340 -0.9632  0.8368 -0.9601  0.8395 -0.9570  0.8422 -0.9538  0.8450 -0.9507  0.8477 -0.9474  0.8505 -0.9442  0.8532 -0.9410  0.8560 -0.9377  0.8587 -0.9344  0.8615 -0.9310  0.8642 -0.9277  0.8669 -0.9242  0.8697 -0.9208  0.8724 -0.9173  0.8752 -0.9138  0.8779 -0.9102  0.8807 -0.9066  0.8834 -0.9030  0.8861 -0.8993  0.8889 -0.8956  0.8916 -0.8919  0.8944 -0.8881  0.8971 -0.8842  0.8999 -0.8803  0.9026 -0.8764  0.9053 -0.8723  0.9081 -0.8683  0.9108 -0.8642  0.9136 -0.8600  0.9163 -0.8557  0.9191 -0.8513  0.9218 -0.8469  0.9246 -0.8425  0.9273 -0.8380  0.9300 -0.8335  0.9328 -0.8288  0.9355 -0.8241  0.9383 -0.8193  0.9410 -0.8144  0.9438 -0.8094  0.9465 -0.8043  0.9492 -0.7991  0.9520 -0.7939  0.9547 -0.7885  0.9575 -0.7831  0.9602 -0.7774  0.9630 -0.7713  0.9657 -0.7653  0.9684 -0.7593  0.9712 -0.7530  0.9739 -0.7465  0.9767 -0.7398  0.9794 -0.7329  0.9822 -0.7257  0.9849 -0.7182  0.9877 -0.7103  0.9904 -0.7022  0.9931 -0.6935  0.9959 -0.6840  0.9986 -0.6744  /

\color{red}
\put {\Large$\bullet$} at 0.1097 -1.3697

\color{black}
\normalcolor

\color{blue}
\put {\LARGE$\bullet$} at 0.0 -1.5442
\colM
\setdashes <7pt>
\setplotsymbol ({\LARGE$\cdot$})
\plot 0.0250 -1.4040  0.0500 -1.3971  0.0750 -1.3900  0.1000 -1.3826  0.1250 -1.3750  0.1500 -1.3672  0.1750 -1.3591  0.2000 -1.3507  0.2250 -1.3420  0.2500 -1.3330  0.2750 -1.3237  0.3000 -1.3140  0.3250 -1.3040  0.3500 -1.2937  0.3750 -1.2829  0.4000 -1.2717  0.4250 -1.2601  0.4500 -1.2480  0.4750 -1.2355  0.5000 -1.2223  0.5250 -1.2087  0.5500 -1.1944  0.5750 -1.1794  0.6000 -1.1638  0.6250 -1.1473  0.6500 -1.1300  0.6750 -1.1118  0.7000 -1.0926  0.7250 -1.0722  0.7500 -1.0506  0.7750 -1.0274  0.8000 -1.0027  0.8250 -0.9760  0.8500 -0.9470  0.8750 -0.9152  0.9000 -0.8799  0.9250 -0.8400  0.9500 -0.7935  0.9750 -0.7364  1.0000 -0.6477   /

\normalcolor
\color{black}
\endpicture

%% file: Figures/figure16.tex
\beginpicture
\color{black}
\setcoordinatesystem units <275pt,135pt>
\axes{0}{-1.7}{0}{0}{1}{0}{-0.245}
\axis left shiftedto x=0 label {\Large$f_t(\phi)$} 
      ticks
      withvalues -0.5 -1.0 /
      at -0.5 -1.0 /
 /  
\axis bottom shiftedto y=0 
    ticks
    withvalues 0.25 0.75 /
    at 0.25 0.75  /
/

\put {\Large$\phi$} at 0.6 -0.12

\color{DarkBlue}
\plot 
0.0060 -1.6130  0.0080 -1.4757  0.0100 -1.4152  0.0120 -1.3845  0.0140 -1.3688  0.0160 -1.3608  0.0180 -1.3576  0.0200 -1.3562 
0.0220 -1.3559  0.0240 -1.3569  0.0260 -1.3583  0.0280 -1.3604  0.0300 -1.3623  0.0320 -1.3647  0.0340 -1.3669  0.0360 -1.3690  0.0380 -1.3713  0.0400 -1.3729  0.0420 -1.3752  0.0440 -1.3767  0.0460 -1.3785  0.0480 -1.3802  0.0500 -1.3815  0.0520 -1.3827  0.0540 -1.3839  0.0560 -1.3850  0.0580 -1.3860  0.0600 -1.3870  0.0620 -1.3880  0.0640 -1.3890  0.0660 -1.3899  0.0680 -1.3905  0.0700 -1.3910  0.0720 -1.3915  0.0740 -1.3918  0.0760 -1.3926  0.0780 -1.3929  0.0800 -1.3931  0.0820 -1.3932  0.0840 -1.3933  0.0860 -1.3934  0.0880 -1.3934  0.0900 -1.3935  0.0920 -1.3935  0.0940 -1.3935  0.0960 -1.3935  0.0980 -1.3932  0.1000 -1.3932  0.1020 -1.3931  0.1040 -1.3930  0.1060 -1.3929  0.1080 -1.3926  0.1100 -1.3924  0.1120 -1.3922  0.1140 -1.3920  0.1160 -1.3917  0.1180 -1.3914  0.1200 -1.3913  0.1220 -1.3909  0.1240 -1.3906  0.1260 -1.3903  0.1280 -1.3901  0.1300 -1.3897  0.1320 -1.3894  0.1340 -1.3890  0.1360 -1.3885  0.1380 -1.3881  0.1400 -1.3878  0.1420 -1.3874  0.1440 -1.3870  0.1460 -1.3864  0.1480 -1.3859  0.1500 -1.3853  0.1520 -1.3849  0.1540 -1.3845  0.1560 -1.3840  0.1580 -1.3835  0.1600 -1.3830  0.1620 -1.3825  0.1640 -1.3821  0.1660 -1.3815  0.1680 -1.3809  0.1700 -1.3805  0.1720 -1.3799  0.1740 -1.3793  0.1760 -1.3787  0.1780 -1.3782  0.1800 -1.3776  0.1820 -1.3770  0.1840 -1.3763  0.1860 -1.3757  0.1880 -1.3750  0.1900 -1.3744  0.1920 -1.3737  0.1940 -1.3731  0.1960 -1.3725  0.1980 -1.3718  0.2000 -1.3712  0.2020 -1.3706  0.2040 -1.3700  0.2060 -1.3693  0.2080 -1.3685  0.2100 -1.3680  0.2120 -1.3673  0.2140 -1.3667  0.2160 -1.3660  0.2180 -1.3653  0.2200 -1.3647  0.2220 -1.3640  0.2240 -1.3633  0.2260 -1.3627  0.2280 -1.3619  0.2300 -1.3612  0.2320 -1.3605  0.2340 -1.3597  0.2360 -1.3591  0.2380 -1.3583  0.2400 -1.3577  0.2420 -1.3571  0.2440 -1.3563  0.2460 -1.3556  0.2480 -1.3549  0.2500 -1.3541  0.2520 -1.3534  0.2540 -1.3527  0.2560 -1.3519  0.2580 -1.3511  0.2600 -1.3504  0.2620 -1.3497  0.2640 -1.3489  0.2660 -1.3482  0.2680 -1.3474  0.2700 -1.3466  0.2720 -1.3458  0.2740 -1.3451  0.2760 -1.3443  0.2780 -1.3435  0.2800 -1.3428  0.2820 -1.3420  0.2840 -1.3412  0.2860 -1.3405  0.2880 -1.3397  0.2900 -1.3389  0.2920 -1.3380  0.2940 -1.3373  0.2960 -1.3365  0.2980 -1.3357  0.3000 -1.3348  0.3020 -1.3340  0.3040 -1.3332  0.3060 -1.3324  0.3080 -1.3316  0.3100 -1.3308  0.3120 -1.3300  0.3140 -1.3291  0.3160 -1.3283  0.3180 -1.3274  0.3200 -1.3266  0.3220 -1.3258  0.3240 -1.3248  0.3260 -1.3240  0.3280 -1.3232  0.3300 -1.3224  0.3320 -1.3215  0.3340 -1.3206  0.3360 -1.3198  0.3380 -1.3189  0.3400 -1.3180  0.3420 -1.3172  0.3440 -1.3164  0.3460 -1.3156  0.3480 -1.3147  0.3500 -1.3138  0.3520 -1.3130  0.3540 -1.3122  0.3560 -1.3113  0.3580 -1.3104  0.3600 -1.3095  0.3620 -1.3085  0.3640 -1.3077  0.3660 -1.3068  0.3680 -1.3060  0.3700 -1.3053  0.3720 -1.3042  0.3740 -1.3035  0.3760 -1.3025  0.3780 -1.3014  0.3800 -1.3008  0.3820 -1.2999  0.3840 -1.2991  0.3860 -1.2980  0.3880 -1.2971  0.3900 -1.2962  0.3920 -1.2953  0.3940 -1.2942  0.3960 -1.2932  0.3980 -1.2923  0.4000 -1.2914  0.4020 -1.2905  0.4040 -1.2895  0.4060 -1.2885  0.4080 -1.2873  0.4100 -1.2865  0.4120 -1.2856  0.4140 -1.2844  0.4160 -1.2838  0.4180 -1.2827  0.4200 -1.2818  0.4220 -1.2809  0.4240 -1.2798  0.4260 -1.2789  0.4280 -1.2779  0.4300 -1.2769  0.4320 -1.2759  0.4340 -1.2747  0.4360 -1.2738  0.4380 -1.2727  0.4400 -1.2717  0.4420 -1.2708  0.4440 -1.2698  0.4460 -1.2687  0.4480 -1.2677  0.4500 -1.2666  0.4520 -1.2656  0.4540 -1.2646  0.4560 -1.2635  0.4580 -1.2625  0.4600 -1.2614  0.4620 -1.2604  0.4640 -1.2593  0.4660 -1.2582  0.4680 -1.2571  0.4700 -1.2561  0.4720 -1.2550  0.4740 -1.2539  0.4760 -1.2528  0.4780 -1.2517  0.4800 -1.2506  0.4820 -1.2495  0.4840 -1.2484  0.4860 -1.2473  0.4880 -1.2462  0.4900 -1.2451  0.4920 -1.2440  0.4940 -1.2429  0.4960 -1.2418  0.4980 -1.2406  0.5000 -1.2395  0.5020 -1.2384  0.5040 -1.2373  0.5060 -1.2361  0.5080 -1.2350  0.5100 -1.2338  0.5120 -1.2327  0.5140 -1.2315  0.5160 -1.2304  0.5180 -1.2292  0.5200 -1.2280  0.5220 -1.2269  0.5240 -1.2257  0.5260 -1.2245  0.5280 -1.2233  0.5300 -1.2222  0.5320 -1.2210  0.5340 -1.2198  0.5360 -1.2186  0.5380 -1.2174  0.5400 -1.2162  0.5420 -1.2150  0.5440 -1.2138  0.5460 -1.2125  0.5480 -1.2113  0.5500 -1.2101  0.5520 -1.2089  0.5540 -1.2077  0.5560 -1.2064  0.5580 -1.2052  0.5600 -1.2040  0.5620 -1.2027  0.5640 -1.2015  0.5660 -1.2002  0.5680 -1.1989  0.5700 -1.1977  0.5720 -1.1964  0.5740 -1.1951  0.5760 -1.1939  0.5780 -1.1926  0.5800 -1.1913  0.5820 -1.1900  0.5840 -1.1887  0.5860 -1.1874  0.5880 -1.1861  0.5900 -1.1848  0.5920 -1.1835  0.5940 -1.1822  0.5960 -1.1809  0.5980 -1.1795  0.6000 -1.1782  0.6020 -1.1769  0.6040 -1.1755  0.6060 -1.1742  0.6080 -1.1728  0.6100 -1.1715  0.6120 -1.1701  0.6140 -1.1688  0.6160 -1.1674  0.6180 -1.1660  0.6200 -1.1647  0.6220 -1.1633  0.6240 -1.1619  0.6260 -1.1605  0.6280 -1.1591  0.6300 -1.1577  0.6320 -1.1563  0.6340 -1.1549  0.6360 -1.1535  0.6380 -1.1521  0.6400 -1.1506  0.6420 -1.1492  0.6440 -1.1477  0.6460 -1.1463  0.6480 -1.1448  0.6500 -1.1434  0.6520 -1.1419  0.6540 -1.1404  0.6560 -1.1390  0.6580 -1.1375  0.6600 -1.1360  0.6620 -1.1345  0.6640 -1.1330  0.6660 -1.1315  0.6680 -1.1300  0.6700 -1.1285  0.6720 -1.1270  0.6740 -1.1254  0.6760 -1.1239  0.6780 -1.1224  0.6800 -1.1208  0.6820 -1.1193  0.6840 -1.1177  0.6860 -1.1162  0.6880 -1.1146  0.6900 -1.1131  0.6920 -1.1115  0.6940 -1.1099  0.6960 -1.1083  0.6980 -1.1067  0.7000 -1.1051  0.7020 -1.1035  0.7040 -1.1019  0.7060 -1.1003  0.7080 -1.0987  0.7100 -1.0970  0.7120 -1.0954  0.7140 -1.0937  0.7160 -1.0921  0.7180 -1.0904  0.7200 -1.0888  0.7220 -1.0871  0.7240 -1.0854  0.7260 -1.0837  0.7280 -1.0820  0.7300 -1.0803  0.7320 -1.0786  0.7340 -1.0769  0.7360 -1.0751  0.7380 -1.0734  0.7400 -1.0716  0.7420 -1.0699  0.7440 -1.0681  0.7460 -1.0663  0.7480 -1.0646  0.7500 -1.0628  0.7520 -1.0610  0.7540 -1.0592  0.7560 -1.0573  0.7580 -1.0555  0.7600 -1.0537  0.7620 -1.0518  0.7640 -1.0500  0.7660 -1.0481  0.7680 -1.0463  0.7700 -1.0444  0.7720 -1.0425  0.7740 -1.0406  0.7760 -1.0387  0.7780 -1.0367  0.7800 -1.0348  0.7820 -1.0329  0.7840 -1.0309  0.7860 -1.0290  0.7880 -1.0270  0.7900 -1.0250  0.7920 -1.0230  0.7940 -1.0210  0.7960 -1.0190  0.7980 -1.0170  0.8000 -1.0149  0.8020 -1.0129  0.8040 -1.0108  0.8060 -1.0088  0.8080 -1.0067  0.8100 -1.0046  0.8120 -1.0025  0.8140 -1.0004  0.8160 -0.9982  0.8180 -0.9961  0.8200 -0.9940  0.8220 -0.9918  0.8240 -0.9897  0.8260 -0.9875  0.8280 -0.9852  0.8300 -0.9830  0.8320 -0.9808  0.8340 -0.9785  0.8360 -0.9763  0.8380 -0.9740  0.8400 -0.9717  0.8420 -0.9695  0.8440 -0.9671  0.8460 -0.9648  0.8480 -0.9625  0.8500 -0.9601  0.8520 -0.9577  0.8540 -0.9553  0.8560 -0.9529  0.8580 -0.9505  0.8600 -0.9480  0.8620 -0.9456  0.8640 -0.9431  0.8660 -0.9406  0.8680 -0.9381  0.8700 -0.9355  0.8720 -0.9330  0.8740 -0.9304  0.8760 -0.9278  0.8780 -0.9252  0.8800 -0.9226  0.8820 -0.9199  0.8840 -0.9173  0.8860 -0.9146  0.8880 -0.9118  0.8900 -0.9091  0.8920 -0.9063  0.8940 -0.9035  0.8960 -0.9008  0.8980 -0.8979  0.9000 -0.8950  0.9020 -0.8922  0.9040 -0.8893  0.9060 -0.8863  0.9080 -0.8834  0.9100 -0.8804  0.9120 -0.8774  0.9140 -0.8743  0.9160 -0.8713  0.9180 -0.8681  0.9200 -0.8646  0.9220 -0.8614  0.9240 -0.8582  0.9260 -0.8550  0.9280 -0.8515  0.9300 -0.8482  0.9320 -0.8448  0.9340 -0.8414  0.9360 -0.8379  0.9380 -0.8344  0.9400 -0.8308  0.9420 -0.8272  0.9440 -0.8235  0.9460 -0.8198  0.9480 -0.8161  0.9500 -0.8123  0.9520 -0.8085  0.9540 -0.8045  0.9560 -0.8004  0.9580 -0.7963  0.9600 -0.7922  0.9620 -0.7880  0.9640 -0.7837  0.9660 -0.7792  0.9680 -0.7747  0.9700 -0.7701  0.9720 -0.7655  0.9740 -0.7608  0.9760 -0.7559  0.9780 -0.7509  0.9800 -0.7458  0.9820 -0.7405  0.9840 -0.7351  0.9860 -0.7296  0.9880 -0.7237  0.9900 -0.7177  0.9920 -0.7114  0.9940 -0.7047  0.9960 -0.6978  0.9980 -0.6902  1.0000 -0.6802   /
\multiput {\large$\cdot$} at 
0.0060 -1.6130  0.0080 -1.4757  0.0100 -1.4152  0.0120 -1.3845  0.0140 -1.3688  0.0160 -1.3608  0.0180 -1.3576  0.0200 -1.3562 
0.0220 -1.3559  0.0240 -1.3569  0.0260 -1.3583  0.0280 -1.3604  0.0300 -1.3623  0.0320 -1.3647  0.0340 -1.3669  0.0360 -1.3690  0.0380 -1.3713  0.0400 -1.3729  0.0420 -1.3752  0.0440 -1.3767  0.0460 -1.3785  0.0480 -1.3802  0.0500 -1.3815  0.0520 -1.3827  0.0540 -1.3839  0.0560 -1.3850  0.0580 -1.3860  0.0600 -1.3870  0.0620 -1.3880  0.0640 -1.3890  0.0660 -1.3899  0.0680 -1.3905  0.0700 -1.3910  0.0720 -1.3915  0.0740 -1.3918  0.0760 -1.3926  0.0780 -1.3929  0.0800 -1.3931  0.0820 -1.3932  0.0840 -1.3933  0.0860 -1.3934  0.0880 -1.3934  0.0900 -1.3935  0.0920 -1.3935  0.0940 -1.3935  0.0960 -1.3935  0.0980 -1.3932  0.1000 -1.3932  0.1020 -1.3931  0.1040 -1.3930  0.1060 -1.3929  0.1080 -1.3926  0.1100 -1.3924  0.1120 -1.3922  0.1140 -1.3920  0.1160 -1.3917  0.1180 -1.3914  0.1200 -1.3913  0.1220 -1.3909  0.1240 -1.3906  0.1260 -1.3903  0.1280 -1.3901  0.1300 -1.3897  0.1320 -1.3894  0.1340 -1.3890  0.1360 -1.3885  0.1380 -1.3881  0.1400 -1.3878  0.1420 -1.3874  0.1440 -1.3870  0.1460 -1.3864  0.1480 -1.3859  0.1500 -1.3853  0.1520 -1.3849  0.1540 -1.3845  0.1560 -1.3840  0.1580 -1.3835  0.1600 -1.3830  0.1620 -1.3825  0.1640 -1.3821  0.1660 -1.3815  0.1680 -1.3809  0.1700 -1.3805  0.1720 -1.3799  0.1740 -1.3793  0.1760 -1.3787  0.1780 -1.3782  0.1800 -1.3776  0.1820 -1.3770  0.1840 -1.3763  0.1860 -1.3757  0.1880 -1.3750  0.1900 -1.3744  0.1920 -1.3737  0.1940 -1.3731  0.1960 -1.3725  0.1980 -1.3718  0.2000 -1.3712  0.2020 -1.3706  0.2040 -1.3700  0.2060 -1.3693  0.2080 -1.3685  0.2100 -1.3680  0.2120 -1.3673  0.2140 -1.3667  0.2160 -1.3660  0.2180 -1.3653  0.2200 -1.3647  0.2220 -1.3640  0.2240 -1.3633  0.2260 -1.3627  0.2280 -1.3619  0.2300 -1.3612  0.2320 -1.3605  0.2340 -1.3597  0.2360 -1.3591  0.2380 -1.3583  0.2400 -1.3577  0.2420 -1.3571  0.2440 -1.3563  0.2460 -1.3556  0.2480 -1.3549  0.2500 -1.3541  0.2520 -1.3534  0.2540 -1.3527  0.2560 -1.3519  0.2580 -1.3511  0.2600 -1.3504  0.2620 -1.3497  0.2640 -1.3489  0.2660 -1.3482  0.2680 -1.3474  0.2700 -1.3466  0.2720 -1.3458  0.2740 -1.3451  0.2760 -1.3443  0.2780 -1.3435  0.2800 -1.3428  0.2820 -1.3420  0.2840 -1.3412  0.2860 -1.3405  0.2880 -1.3397  0.2900 -1.3389  0.2920 -1.3380  0.2940 -1.3373  0.2960 -1.3365  0.2980 -1.3357  0.3000 -1.3348  0.3020 -1.3340  0.3040 -1.3332  0.3060 -1.3324  0.3080 -1.3316  0.3100 -1.3308  0.3120 -1.3300  0.3140 -1.3291  0.3160 -1.3283  0.3180 -1.3274  0.3200 -1.3266  0.3220 -1.3258  0.3240 -1.3248  0.3260 -1.3240  0.3280 -1.3232  0.3300 -1.3224  0.3320 -1.3215  0.3340 -1.3206  0.3360 -1.3198  0.3380 -1.3189  0.3400 -1.3180  0.3420 -1.3172  0.3440 -1.3164  0.3460 -1.3156  0.3480 -1.3147  0.3500 -1.3138  0.3520 -1.3130  0.3540 -1.3122  0.3560 -1.3113  0.3580 -1.3104  0.3600 -1.3095  0.3620 -1.3085  0.3640 -1.3077  0.3660 -1.3068  0.3680 -1.3060  0.3700 -1.3053  0.3720 -1.3042  0.3740 -1.3035  0.3760 -1.3025  0.3780 -1.3014  0.3800 -1.3008  0.3820 -1.2999  0.3840 -1.2991  0.3860 -1.2980  0.3880 -1.2971  0.3900 -1.2962  0.3920 -1.2953  0.3940 -1.2942  0.3960 -1.2932  0.3980 -1.2923  0.4000 -1.2914  0.4020 -1.2905  0.4040 -1.2895  0.4060 -1.2885  0.4080 -1.2873  0.4100 -1.2865  0.4120 -1.2856  0.4140 -1.2844  0.4160 -1.2838  0.4180 -1.2827  0.4200 -1.2818  0.4220 -1.2809  0.4240 -1.2798  0.4260 -1.2789  0.4280 -1.2779  0.4300 -1.2769  0.4320 -1.2759  0.4340 -1.2747  0.4360 -1.2738  0.4380 -1.2727  0.4400 -1.2717  0.4420 -1.2708  0.4440 -1.2698  0.4460 -1.2687  0.4480 -1.2677  0.4500 -1.2666  0.4520 -1.2656  0.4540 -1.2646  0.4560 -1.2635  0.4580 -1.2625  0.4600 -1.2614  0.4620 -1.2604  0.4640 -1.2593  0.4660 -1.2582  0.4680 -1.2571  0.4700 -1.2561  0.4720 -1.2550  0.4740 -1.2539  0.4760 -1.2528  0.4780 -1.2517  0.4800 -1.2506  0.4820 -1.2495  0.4840 -1.2484  0.4860 -1.2473  0.4880 -1.2462  0.4900 -1.2451  0.4920 -1.2440  0.4940 -1.2429  0.4960 -1.2418  0.4980 -1.2406  0.5000 -1.2395  0.5020 -1.2384  0.5040 -1.2373  0.5060 -1.2361  0.5080 -1.2350  0.5100 -1.2338  0.5120 -1.2327  0.5140 -1.2315  0.5160 -1.2304  0.5180 -1.2292  0.5200 -1.2280  0.5220 -1.2269  0.5240 -1.2257  0.5260 -1.2245  0.5280 -1.2233  0.5300 -1.2222  0.5320 -1.2210  0.5340 -1.2198  0.5360 -1.2186  0.5380 -1.2174  0.5400 -1.2162  0.5420 -1.2150  0.5440 -1.2138  0.5460 -1.2125  0.5480 -1.2113  0.5500 -1.2101  0.5520 -1.2089  0.5540 -1.2077  0.5560 -1.2064  0.5580 -1.2052  0.5600 -1.2040  0.5620 -1.2027  0.5640 -1.2015  0.5660 -1.2002  0.5680 -1.1989  0.5700 -1.1977  0.5720 -1.1964  0.5740 -1.1951  0.5760 -1.1939  0.5780 -1.1926  0.5800 -1.1913  0.5820 -1.1900  0.5840 -1.1887  0.5860 -1.1874  0.5880 -1.1861  0.5900 -1.1848  0.5920 -1.1835  0.5940 -1.1822  0.5960 -1.1809  0.5980 -1.1795  0.6000 -1.1782  0.6020 -1.1769  0.6040 -1.1755  0.6060 -1.1742  0.6080 -1.1728  0.6100 -1.1715  0.6120 -1.1701  0.6140 -1.1688  0.6160 -1.1674  0.6180 -1.1660  0.6200 -1.1647  0.6220 -1.1633  0.6240 -1.1619  0.6260 -1.1605  0.6280 -1.1591  0.6300 -1.1577  0.6320 -1.1563  0.6340 -1.1549  0.6360 -1.1535  0.6380 -1.1521  0.6400 -1.1506  0.6420 -1.1492  0.6440 -1.1477  0.6460 -1.1463  0.6480 -1.1448  0.6500 -1.1434  0.6520 -1.1419  0.6540 -1.1404  0.6560 -1.1390  0.6580 -1.1375  0.6600 -1.1360  0.6620 -1.1345  0.6640 -1.1330  0.6660 -1.1315  0.6680 -1.1300  0.6700 -1.1285  0.6720 -1.1270  0.6740 -1.1254  0.6760 -1.1239  0.6780 -1.1224  0.6800 -1.1208  0.6820 -1.1193  0.6840 -1.1177  0.6860 -1.1162  0.6880 -1.1146  0.6900 -1.1131  0.6920 -1.1115  0.6940 -1.1099  0.6960 -1.1083  0.6980 -1.1067  0.7000 -1.1051  0.7020 -1.1035  0.7040 -1.1019  0.7060 -1.1003  0.7080 -1.0987  0.7100 -1.0970  0.7120 -1.0954  0.7140 -1.0937  0.7160 -1.0921  0.7180 -1.0904  0.7200 -1.0888  0.7220 -1.0871  0.7240 -1.0854  0.7260 -1.0837  0.7280 -1.0820  0.7300 -1.0803  0.7320 -1.0786  0.7340 -1.0769  0.7360 -1.0751  0.7380 -1.0734  0.7400 -1.0716  0.7420 -1.0699  0.7440 -1.0681  0.7460 -1.0663  0.7480 -1.0646  0.7500 -1.0628  0.7520 -1.0610  0.7540 -1.0592  0.7560 -1.0573  0.7580 -1.0555  0.7600 -1.0537  0.7620 -1.0518  0.7640 -1.0500  0.7660 -1.0481  0.7680 -1.0463  0.7700 -1.0444  0.7720 -1.0425  0.7740 -1.0406  0.7760 -1.0387  0.7780 -1.0367  0.7800 -1.0348  0.7820 -1.0329  0.7840 -1.0309  0.7860 -1.0290  0.7880 -1.0270  0.7900 -1.0250  0.7920 -1.0230  0.7940 -1.0210  0.7960 -1.0190  0.7980 -1.0170  0.8000 -1.0149  0.8020 -1.0129  0.8040 -1.0108  0.8060 -1.0088  0.8080 -1.0067  0.8100 -1.0046  0.8120 -1.0025  0.8140 -1.0004  0.8160 -0.9982  0.8180 -0.9961  0.8200 -0.9940  0.8220 -0.9918  0.8240 -0.9897  0.8260 -0.9875  0.8280 -0.9852  0.8300 -0.9830  0.8320 -0.9808  0.8340 -0.9785  0.8360 -0.9763  0.8380 -0.9740  0.8400 -0.9717  0.8420 -0.9695  0.8440 -0.9671  0.8460 -0.9648  0.8480 -0.9625  0.8500 -0.9601  0.8520 -0.9577  0.8540 -0.9553  0.8560 -0.9529  0.8580 -0.9505  0.8600 -0.9480  0.8620 -0.9456  0.8640 -0.9431  0.8660 -0.9406  0.8680 -0.9381  0.8700 -0.9355  0.8720 -0.9330  0.8740 -0.9304  0.8760 -0.9278  0.8780 -0.9252  0.8800 -0.9226  0.8820 -0.9199  0.8840 -0.9173  0.8860 -0.9146  0.8880 -0.9118  0.8900 -0.9091  0.8920 -0.9063  0.8940 -0.9035  0.8960 -0.9008  0.8980 -0.8979  0.9000 -0.8950  0.9020 -0.8922  0.9040 -0.8893  0.9060 -0.8863  0.9080 -0.8834  0.9100 -0.8804  0.9120 -0.8774  0.9140 -0.8743  0.9160 -0.8713  0.9180 -0.8681  0.9200 -0.8646  0.9220 -0.8614  0.9240 -0.8582  0.9260 -0.8550  0.9280 -0.8515  0.9300 -0.8482  0.9320 -0.8448  0.9340 -0.8414  0.9360 -0.8379  0.9380 -0.8344  0.9400 -0.8308  0.9420 -0.8272  0.9440 -0.8235  0.9460 -0.8198  0.9480 -0.8161  0.9500 -0.8123  0.9520 -0.8085  0.9540 -0.8045  0.9560 -0.8004  0.9580 -0.7963  0.9600 -0.7922  0.9620 -0.7880  0.9640 -0.7837  0.9660 -0.7792  0.9680 -0.7747  0.9700 -0.7701  0.9720 -0.7655  0.9740 -0.7608  0.9760 -0.7559  0.9780 -0.7509  0.9800 -0.7458  0.9820 -0.7405  0.9840 -0.7351  0.9860 -0.7296  0.9880 -0.7237  0.9900 -0.7177  0.9920 -0.7114  0.9940 -0.7047  0.9960 -0.6978  0.9980 -0.6902  1.0000 -0.6802  /

\color{DarkGreen}
\plot 0.0240 -0.5827  0.0260 -0.7053  0.0280 -0.7916  0.0300 -0.8584  0.0320 -0.9123  0.0340 -0.9570  0.0360 -0.9949  0.0380 -1.0270  0.0400 -1.0551  0.0420 -1.0798  0.0440 -1.1017  0.0460 -1.1212  0.0480 -1.1393  0.0500 -1.1550  0.0520 -1.1700  0.0540 -1.1829  0.0560 -1.1949  0.0580 -1.2058  0.0600 -1.2159  0.0620 -1.2252  0.0640 -1.2337  0.0660 -1.2417  0.0680 -1.2488  0.0700 -1.2558  0.0720 -1.2624  0.0740 -1.2684  0.0760 -1.2739  0.0780 -1.2793  0.0800 -1.2844  0.0820 -1.2889  0.0840 -1.2934  0.0860 -1.2974  0.0880 -1.3010  0.0900 -1.3046  0.0920 -1.3080  0.0940 -1.3109  0.0960 -1.3138  0.0980 -1.3165  0.1000 -1.3192  0.1020 -1.3216  0.1040 -1.3240  0.1060 -1.3263  0.1080 -1.3284  0.1100 -1.3303  0.1120 -1.3321  0.1140 -1.3338  0.1160 -1.3351  0.1180 -1.3366  0.1200 -1.3381  0.1220 -1.3393  0.1240 -1.3406  0.1260 -1.3417  0.1280 -1.3427  0.1300 -1.3435  0.1320 -1.3443  0.1340 -1.3452  0.1360 -1.3460  0.1380 -1.3468  0.1400 -1.3474  0.1420 -1.3481  0.1440 -1.3485  0.1460 -1.3491  0.1480 -1.3495  0.1500 -1.3499  0.1520 -1.3504  0.1540 -1.3508  0.1560 -1.3512  0.1580 -1.3515  0.1600 -1.3518  0.1620 -1.3519  0.1640 -1.3521  0.1660 -1.3523  0.1680 -1.3524  0.1700 -1.3524  0.1720 -1.3525  0.1740 -1.3526  0.1760 -1.3526  0.1780 -1.3526  0.1800 -1.3526  0.1820 -1.3525  0.1840 -1.3524  0.1860 -1.3523  0.1880 -1.3523  0.1900 -1.3521  0.1920 -1.3519  0.1940 -1.3518  0.1960 -1.3516  0.1980 -1.3514  0.2000 -1.3513  0.2020 -1.3510  0.2040 -1.3508  0.2060 -1.3506  0.2080 -1.3502  0.2100 -1.3500  0.2120 -1.3497  0.2140 -1.3494  0.2160 -1.3491  0.2180 -1.3487  0.2200 -1.3484  0.2220 -1.3479  0.2240 -1.3476  0.2260 -1.3471  0.2280 -1.3467  0.2300 -1.3462  0.2320 -1.3458  0.2340 -1.3454  0.2360 -1.3449  0.2380 -1.3445  0.2400 -1.3440  0.2420 -1.3435  0.2440 -1.3429  0.2460 -1.3424  0.2480 -1.3419  0.2500 -1.3414  0.2520 -1.3409  0.2540 -1.3404  0.2560 -1.3398  0.2580 -1.3393  0.2600 -1.3387  0.2620 -1.3382  0.2640 -1.3376  0.2660 -1.3370  0.2680 -1.3364  0.2700 -1.3358  0.2720 -1.3351  0.2740 -1.3347  0.2760 -1.3341  0.2780 -1.3334  0.2800 -1.3328  0.2820 -1.3322  0.2840 -1.3315  0.2860 -1.3309  0.2880 -1.3302  0.2900 -1.3296  0.2920 -1.3290  0.2940 -1.3283  0.2960 -1.3277  0.2980 -1.3270  0.3000 -1.3263  0.3020 -1.3255  0.3040 -1.3248  0.3060 -1.3241  0.3080 -1.3235  0.3100 -1.3227  0.3120 -1.3220  0.3140 -1.3213  0.3160 -1.3205  0.3180 -1.3198  0.3200 -1.3191  0.3220 -1.3183  0.3240 -1.3176  0.3260 -1.3168  0.3280 -1.3162  0.3300 -1.3153  0.3320 -1.3146  0.3340 -1.3139  0.3360 -1.3132  0.3380 -1.3124  0.3400 -1.3117  0.3420 -1.3109  0.3440 -1.3101  0.3460 -1.3094  0.3480 -1.3085  0.3500 -1.3078  0.3520 -1.3070  0.3540 -1.3063  0.3560 -1.3054  0.3580 -1.3047  0.3600 -1.3039  0.3620 -1.3032  0.3640 -1.3024  0.3660 -1.3015  0.3680 -1.3008  0.3700 -1.2998  0.3720 -1.2992  0.3740 -1.2982  0.3760 -1.2974  0.3780 -1.2967  0.3800 -1.2957  0.3820 -1.2951  0.3840 -1.2943  0.3860 -1.2935  0.3880 -1.2925  0.3900 -1.2915  0.3920 -1.2908  0.3940 -1.2899  0.3960 -1.2890  0.3980 -1.2884  0.4000 -1.2875  0.4020 -1.2866  0.4040 -1.2857  0.4060 -1.2848  0.4080 -1.2841  0.4100 -1.2831  0.4120 -1.2822  0.4140 -1.2812  0.4160 -1.2803  0.4180 -1.2794  0.4200 -1.2784  0.4220 -1.2773  0.4240 -1.2765  0.4260 -1.2755  0.4280 -1.2744  0.4300 -1.2736  0.4320 -1.2726  0.4340 -1.2716  0.4360 -1.2706  0.4380 -1.2697  0.4400 -1.2687  0.4420 -1.2677  0.4440 -1.2666  0.4460 -1.2657  0.4480 -1.2647  0.4500 -1.2637  0.4520 -1.2627  0.4540 -1.2616  0.4560 -1.2606  0.4580 -1.2596  0.4600 -1.2586  0.4620 -1.2575  0.4640 -1.2564  0.4660 -1.2555  0.4680 -1.2545  0.4700 -1.2534  0.4720 -1.2524  0.4740 -1.2513  0.4760 -1.2503  0.4780 -1.2492  0.4800 -1.2482  0.4820 -1.2471  0.4840 -1.2460  0.4860 -1.2450  0.4880 -1.2439  0.4900 -1.2428  0.4920 -1.2417  0.4940 -1.2406  0.4960 -1.2395  0.4980 -1.2384  0.5000 -1.2373  0.5020 -1.2362  0.5040 -1.2351  0.5060 -1.2340  0.5080 -1.2329  0.5100 -1.2317  0.5120 -1.2306  0.5140 -1.2294  0.5160 -1.2283  0.5180 -1.2272  0.5200 -1.2260  0.5220 -1.2249  0.5240 -1.2238  0.5260 -1.2226  0.5280 -1.2214  0.5300 -1.2203  0.5320 -1.2191  0.5340 -1.2179  0.5360 -1.2167  0.5380 -1.2156  0.5400 -1.2144  0.5420 -1.2132  0.5440 -1.2120  0.5460 -1.2108  0.5480 -1.2096  0.5500 -1.2084  0.5520 -1.2072  0.5540 -1.2060  0.5560 -1.2048  0.5580 -1.2036  0.5600 -1.2023  0.5620 -1.2011  0.5640 -1.1999  0.5660 -1.1986  0.5680 -1.1974  0.5700 -1.1961  0.5720 -1.1949  0.5740 -1.1936  0.5760 -1.1924  0.5780 -1.1911  0.5800 -1.1899  0.5820 -1.1886  0.5840 -1.1873  0.5860 -1.1860  0.5880 -1.1847  0.5900 -1.1835  0.5920 -1.1822  0.5940 -1.1808  0.5960 -1.1795  0.5980 -1.1782  0.6000 -1.1769  0.6020 -1.1756  0.6040 -1.1743  0.6060 -1.1729  0.6080 -1.1716  0.6100 -1.1703  0.6120 -1.1689  0.6140 -1.1675  0.6160 -1.1662  0.6180 -1.1648  0.6200 -1.1635  0.6220 -1.1621  0.6240 -1.1607  0.6260 -1.1593  0.6280 -1.1580  0.6300 -1.1566  0.6320 -1.1552  0.6340 -1.1538  0.6360 -1.1524  0.6380 -1.1510  0.6400 -1.1495  0.6420 -1.1481  0.6440 -1.1467  0.6460 -1.1452  0.6480 -1.1438  0.6500 -1.1423  0.6520 -1.1409  0.6540 -1.1394  0.6560 -1.1379  0.6580 -1.1365  0.6600 -1.1350  0.6620 -1.1335  0.6640 -1.1321  0.6660 -1.1306  0.6680 -1.1291  0.6700 -1.1276  0.6720 -1.1261  0.6740 -1.1245  0.6760 -1.1230  0.6780 -1.1215  0.6800 -1.1199  0.6820 -1.1184  0.6840 -1.1168  0.6860 -1.1153  0.6880 -1.1137  0.6900 -1.1122  0.6920 -1.1106  0.6940 -1.1090  0.6960 -1.1075  0.6980 -1.1059  0.7000 -1.1043  0.7020 -1.1027  0.7040 -1.1011  0.7060 -1.0995  0.7080 -1.0979  0.7100 -1.0962  0.7120 -1.0946  0.7140 -1.0929  0.7160 -1.0913  0.7180 -1.0896  0.7200 -1.0880  0.7220 -1.0863  0.7240 -1.0846  0.7260 -1.0829  0.7280 -1.0812  0.7300 -1.0795  0.7320 -1.0778  0.7340 -1.0761  0.7360 -1.0744  0.7380 -1.0727  0.7400 -1.0709  0.7420 -1.0692  0.7440 -1.0674  0.7460 -1.0656  0.7480 -1.0639  0.7500 -1.0621  0.7520 -1.0603  0.7540 -1.0585  0.7560 -1.0567  0.7580 -1.0548  0.7600 -1.0530  0.7620 -1.0512  0.7640 -1.0493  0.7660 -1.0475  0.7680 -1.0456  0.7700 -1.0437  0.7720 -1.0419  0.7740 -1.0400  0.7760 -1.0381  0.7780 -1.0361  0.7800 -1.0342  0.7820 -1.0323  0.7840 -1.0303  0.7860 -1.0284  0.7880 -1.0264  0.7900 -1.0244  0.7920 -1.0224  0.7940 -1.0204  0.7960 -1.0184  0.7980 -1.0164  0.8000 -1.0144  0.8020 -1.0123  0.8040 -1.0103  0.8060 -1.0082  0.8080 -1.0061  0.8100 -1.0040  0.8120 -1.0019  0.8140 -0.9998  0.8160 -0.9977  0.8180 -0.9956  0.8200 -0.9934  0.8220 -0.9912  0.8240 -0.9890  0.8260 -0.9869  0.8280 -0.9846  0.8300 -0.9824  0.8320 -0.9802  0.8340 -0.9779  0.8360 -0.9757  0.8380 -0.9734  0.8400 -0.9711  0.8420 -0.9688  0.8440 -0.9665  0.8460 -0.9642  0.8480 -0.9618  0.8500 -0.9594  0.8520 -0.9571  0.8540 -0.9547  0.8560 -0.9523  0.8580 -0.9498  0.8600 -0.9474  0.8620 -0.9449  0.8640 -0.9424  0.8660 -0.9399  0.8680 -0.9374  0.8700 -0.9349  0.8720 -0.9323  0.8740 -0.9298  0.8760 -0.9272  0.8780 -0.9245  0.8800 -0.9219  0.8820 -0.9193  0.8840 -0.9166  0.8860 -0.9139  0.8880 -0.9112  0.8900 -0.9084  0.8920 -0.9056  0.8940 -0.9028  0.8960 -0.9000  0.8980 -0.8972  0.9000 -0.8943  0.9020 -0.8914  0.9040 -0.8885  0.9060 -0.8855  0.9080 -0.8826  0.9100 -0.8796  0.9120 -0.8765  0.9140 -0.8735  0.9160 -0.8703  0.9180 -0.8672  0.9200 -0.8641  0.9220 -0.8609  0.9240 -0.8576  0.9260 -0.8544  0.9280 -0.8511  0.9300 -0.8477  0.9320 -0.8444  0.9340 -0.8410  0.9360 -0.8375  0.9380 -0.8340  0.9400 -0.8305  0.9420 -0.8269  0.9440 -0.8233  0.9460 -0.8195  0.9480 -0.8158  0.9500 -0.8120  0.9520 -0.8082  0.9540 -0.8043  0.9560 -0.8003  0.9580 -0.7961  0.9600 -0.7920  0.9620 -0.7878  0.9640 -0.7834  0.9660 -0.7791  0.9680 -0.7746  0.9700 -0.7701  0.9720 -0.7654  0.9740 -0.7606  0.9760 -0.7558  0.9780 -0.7508  0.9800 -0.7455  0.9820 -0.7403  0.9840 -0.7348  0.9860 -0.7292  0.9880 -0.7233  0.9900 -0.7173  0.9920 -0.7110  0.9940 -0.7045  0.9960 -0.6974  0.9980 -0.6903  1.0000 -0.6799     /
\multiput {\large$\cdot$} at 0.0240 -0.5827  0.0260 -0.7053  0.0280 -0.7916  0.0300 -0.8584  0.0320 -0.9123  0.0340 -0.9570  0.0360 -0.9949  0.0380 -1.0270  0.0400 -1.0551  0.0420 -1.0798  0.0440 -1.1017  0.0460 -1.1212  0.0480 -1.1393  0.0500 -1.1550  0.0520 -1.1700  0.0540 -1.1829  0.0560 -1.1949  0.0580 -1.2058  0.0600 -1.2159  0.0620 -1.2252  0.0640 -1.2337  0.0660 -1.2417  0.0680 -1.2488  0.0700 -1.2558  0.0720 -1.2624  0.0740 -1.2684  0.0760 -1.2739  0.0780 -1.2793  0.0800 -1.2844  0.0820 -1.2889  0.0840 -1.2934  0.0860 -1.2974  0.0880 -1.3010  0.0900 -1.3046  0.0920 -1.3080  0.0940 -1.3109  0.0960 -1.3138  0.0980 -1.3165  0.1000 -1.3192  0.1020 -1.3216  0.1040 -1.3240  0.1060 -1.3263  0.1080 -1.3284  0.1100 -1.3303  0.1120 -1.3321  0.1140 -1.3338  0.1160 -1.3351  0.1180 -1.3366  0.1200 -1.3381  0.1220 -1.3393  0.1240 -1.3406  0.1260 -1.3417  0.1280 -1.3427  0.1300 -1.3435  0.1320 -1.3443  0.1340 -1.3452  0.1360 -1.3460  0.1380 -1.3468  0.1400 -1.3474  0.1420 -1.3481  0.1440 -1.3485  0.1460 -1.3491  0.1480 -1.3495  0.1500 -1.3499  0.1520 -1.3504  0.1540 -1.3508  0.1560 -1.3512  0.1580 -1.3515  0.1600 -1.3518  0.1620 -1.3519  0.1640 -1.3521  0.1660 -1.3523  0.1680 -1.3524  0.1700 -1.3524  0.1720 -1.3525  0.1740 -1.3526  0.1760 -1.3526  0.1780 -1.3526  0.1800 -1.3526  0.1820 -1.3525  0.1840 -1.3524  0.1860 -1.3523  0.1880 -1.3523  0.1900 -1.3521  0.1920 -1.3519  0.1940 -1.3518  0.1960 -1.3516  0.1980 -1.3514  0.2000 -1.3513  0.2020 -1.3510  0.2040 -1.3508  0.2060 -1.3506  0.2080 -1.3502  0.2100 -1.3500  0.2120 -1.3497  0.2140 -1.3494  0.2160 -1.3491  0.2180 -1.3487  0.2200 -1.3484  0.2220 -1.3479  0.2240 -1.3476  0.2260 -1.3471  0.2280 -1.3467  0.2300 -1.3462  0.2320 -1.3458  0.2340 -1.3454  0.2360 -1.3449  0.2380 -1.3445  0.2400 -1.3440  0.2420 -1.3435  0.2440 -1.3429  0.2460 -1.3424  0.2480 -1.3419  0.2500 -1.3414  0.2520 -1.3409  0.2540 -1.3404  0.2560 -1.3398  0.2580 -1.3393  0.2600 -1.3387  0.2620 -1.3382  0.2640 -1.3376  0.2660 -1.3370  0.2680 -1.3364  0.2700 -1.3358  0.2720 -1.3351  0.2740 -1.3347  0.2760 -1.3341  0.2780 -1.3334  0.2800 -1.3328  0.2820 -1.3322  0.2840 -1.3315  0.2860 -1.3309  0.2880 -1.3302  0.2900 -1.3296  0.2920 -1.3290  0.2940 -1.3283  0.2960 -1.3277  0.2980 -1.3270  0.3000 -1.3263  0.3020 -1.3255  0.3040 -1.3248  0.3060 -1.3241  0.3080 -1.3235  0.3100 -1.3227  0.3120 -1.3220  0.3140 -1.3213  0.3160 -1.3205  0.3180 -1.3198  0.3200 -1.3191  0.3220 -1.3183  0.3240 -1.3176  0.3260 -1.3168  0.3280 -1.3162  0.3300 -1.3153  0.3320 -1.3146  0.3340 -1.3139  0.3360 -1.3132  0.3380 -1.3124  0.3400 -1.3117  0.3420 -1.3109  0.3440 -1.3101  0.3460 -1.3094  0.3480 -1.3085  0.3500 -1.3078  0.3520 -1.3070  0.3540 -1.3063  0.3560 -1.3054  0.3580 -1.3047  0.3600 -1.3039  0.3620 -1.3032  0.3640 -1.3024  0.3660 -1.3015  0.3680 -1.3008  0.3700 -1.2998  0.3720 -1.2992  0.3740 -1.2982  0.3760 -1.2974  0.3780 -1.2967  0.3800 -1.2957  0.3820 -1.2951  0.3840 -1.2943  0.3860 -1.2935  0.3880 -1.2925  0.3900 -1.2915  0.3920 -1.2908  0.3940 -1.2899  0.3960 -1.2890  0.3980 -1.2884  0.4000 -1.2875  0.4020 -1.2866  0.4040 -1.2857  0.4060 -1.2848  0.4080 -1.2841  0.4100 -1.2831  0.4120 -1.2822  0.4140 -1.2812  0.4160 -1.2803  0.4180 -1.2794  0.4200 -1.2784  0.4220 -1.2773  0.4240 -1.2765  0.4260 -1.2755  0.4280 -1.2744  0.4300 -1.2736  0.4320 -1.2726  0.4340 -1.2716  0.4360 -1.2706  0.4380 -1.2697  0.4400 -1.2687  0.4420 -1.2677  0.4440 -1.2666  0.4460 -1.2657  0.4480 -1.2647  0.4500 -1.2637  0.4520 -1.2627  0.4540 -1.2616  0.4560 -1.2606  0.4580 -1.2596  0.4600 -1.2586  0.4620 -1.2575  0.4640 -1.2564  0.4660 -1.2555  0.4680 -1.2545  0.4700 -1.2534  0.4720 -1.2524  0.4740 -1.2513  0.4760 -1.2503  0.4780 -1.2492  0.4800 -1.2482  0.4820 -1.2471  0.4840 -1.2460  0.4860 -1.2450  0.4880 -1.2439  0.4900 -1.2428  0.4920 -1.2417  0.4940 -1.2406  0.4960 -1.2395  0.4980 -1.2384  0.5000 -1.2373  0.5020 -1.2362  0.5040 -1.2351  0.5060 -1.2340  0.5080 -1.2329  0.5100 -1.2317  0.5120 -1.2306  0.5140 -1.2294  0.5160 -1.2283  0.5180 -1.2272  0.5200 -1.2260  0.5220 -1.2249  0.5240 -1.2238  0.5260 -1.2226  0.5280 -1.2214  0.5300 -1.2203  0.5320 -1.2191  0.5340 -1.2179  0.5360 -1.2167  0.5380 -1.2156  0.5400 -1.2144  0.5420 -1.2132  0.5440 -1.2120  0.5460 -1.2108  0.5480 -1.2096  0.5500 -1.2084  0.5520 -1.2072  0.5540 -1.2060  0.5560 -1.2048  0.5580 -1.2036  0.5600 -1.2023  0.5620 -1.2011  0.5640 -1.1999  0.5660 -1.1986  0.5680 -1.1974  0.5700 -1.1961  0.5720 -1.1949  0.5740 -1.1936  0.5760 -1.1924  0.5780 -1.1911  0.5800 -1.1899  0.5820 -1.1886  0.5840 -1.1873  0.5860 -1.1860  0.5880 -1.1847  0.5900 -1.1835  0.5920 -1.1822  0.5940 -1.1808  0.5960 -1.1795  0.5980 -1.1782  0.6000 -1.1769  0.6020 -1.1756  0.6040 -1.1743  0.6060 -1.1729  0.6080 -1.1716  0.6100 -1.1703  0.6120 -1.1689  0.6140 -1.1675  0.6160 -1.1662  0.6180 -1.1648  0.6200 -1.1635  0.6220 -1.1621  0.6240 -1.1607  0.6260 -1.1593  0.6280 -1.1580  0.6300 -1.1566  0.6320 -1.1552  0.6340 -1.1538  0.6360 -1.1524  0.6380 -1.1510  0.6400 -1.1495  0.6420 -1.1481  0.6440 -1.1467  0.6460 -1.1452  0.6480 -1.1438  0.6500 -1.1423  0.6520 -1.1409  0.6540 -1.1394  0.6560 -1.1379  0.6580 -1.1365  0.6600 -1.1350  0.6620 -1.1335  0.6640 -1.1321  0.6660 -1.1306  0.6680 -1.1291  0.6700 -1.1276  0.6720 -1.1261  0.6740 -1.1245  0.6760 -1.1230  0.6780 -1.1215  0.6800 -1.1199  0.6820 -1.1184  0.6840 -1.1168  0.6860 -1.1153  0.6880 -1.1137  0.6900 -1.1122  0.6920 -1.1106  0.6940 -1.1090  0.6960 -1.1075  0.6980 -1.1059  0.7000 -1.1043  0.7020 -1.1027  0.7040 -1.1011  0.7060 -1.0995  0.7080 -1.0979  0.7100 -1.0962  0.7120 -1.0946  0.7140 -1.0929  0.7160 -1.0913  0.7180 -1.0896  0.7200 -1.0880  0.7220 -1.0863  0.7240 -1.0846  0.7260 -1.0829  0.7280 -1.0812  0.7300 -1.0795  0.7320 -1.0778  0.7340 -1.0761  0.7360 -1.0744  0.7380 -1.0727  0.7400 -1.0709  0.7420 -1.0692  0.7440 -1.0674  0.7460 -1.0656  0.7480 -1.0639  0.7500 -1.0621  0.7520 -1.0603  0.7540 -1.0585  0.7560 -1.0567  0.7580 -1.0548  0.7600 -1.0530  0.7620 -1.0512  0.7640 -1.0493  0.7660 -1.0475  0.7680 -1.0456  0.7700 -1.0437  0.7720 -1.0419  0.7740 -1.0400  0.7760 -1.0381  0.7780 -1.0361  0.7800 -1.0342  0.7820 -1.0323  0.7840 -1.0303  0.7860 -1.0284  0.7880 -1.0264  0.7900 -1.0244  0.7920 -1.0224  0.7940 -1.0204  0.7960 -1.0184  0.7980 -1.0164  0.8000 -1.0144  0.8020 -1.0123  0.8040 -1.0103  0.8060 -1.0082  0.8080 -1.0061  0.8100 -1.0040  0.8120 -1.0019  0.8140 -0.9998  0.8160 -0.9977  0.8180 -0.9956  0.8200 -0.9934  0.8220 -0.9912  0.8240 -0.9890  0.8260 -0.9869  0.8280 -0.9846  0.8300 -0.9824  0.8320 -0.9802  0.8340 -0.9779  0.8360 -0.9757  0.8380 -0.9734  0.8400 -0.9711  0.8420 -0.9688  0.8440 -0.9665  0.8460 -0.9642  0.8480 -0.9618  0.8500 -0.9594  0.8520 -0.9571  0.8540 -0.9547  0.8560 -0.9523  0.8580 -0.9498  0.8600 -0.9474  0.8620 -0.9449  0.8640 -0.9424  0.8660 -0.9399  0.8680 -0.9374  0.8700 -0.9349  0.8720 -0.9323  0.8740 -0.9298  0.8760 -0.9272  0.8780 -0.9245  0.8800 -0.9219  0.8820 -0.9193  0.8840 -0.9166  0.8860 -0.9139  0.8880 -0.9112  0.8900 -0.9084  0.8920 -0.9056  0.8940 -0.9028  0.8960 -0.9000  0.8980 -0.8972  0.9000 -0.8943  0.9020 -0.8914  0.9040 -0.8885  0.9060 -0.8855  0.9080 -0.8826  0.9100 -0.8796  0.9120 -0.8765  0.9140 -0.8735  0.9160 -0.8703  0.9180 -0.8672  0.9200 -0.8641  0.9220 -0.8609  0.9240 -0.8576  0.9260 -0.8544  0.9280 -0.8511  0.9300 -0.8477  0.9320 -0.8444  0.9340 -0.8410  0.9360 -0.8375  0.9380 -0.8340  0.9400 -0.8305  0.9420 -0.8269  0.9440 -0.8233  0.9460 -0.8195  0.9480 -0.8158  0.9500 -0.8120  0.9520 -0.8082  0.9540 -0.8043  0.9560 -0.8003  0.9580 -0.7961  0.9600 -0.7920  0.9620 -0.7878  0.9640 -0.7834  0.9660 -0.7791  0.9680 -0.7746  0.9700 -0.7701  0.9720 -0.7654  0.9740 -0.7606  0.9760 -0.7558  0.9780 -0.7508  0.9800 -0.7455  0.9820 -0.7403  0.9840 -0.7348  0.9860 -0.7292  0.9880 -0.7233  0.9900 -0.7173  0.9920 -0.7110  0.9940 -0.7045  0.9960 -0.6974  0.9980 -0.6903  1.0000 -0.6799    /

\color{DarkRed}
\plot 0.0300 -0.4620  0.0320 -0.5839  0.0340 -0.6739  0.0360 -0.7452  0.0380 -0.8041  0.0400 -0.8536  0.0420 -0.8960  0.0440 -0.9328  0.0460 -0.9652  0.0480 -0.9937  0.0500 -1.0191  0.0520 -1.0422  0.0540 -1.0628  0.0560 -1.0817  0.0580 -1.0987  0.0600 -1.1144  0.0620 -1.1287  0.0640 -1.1420  0.0660 -1.1544  0.0680 -1.1657  0.0700 -1.1762  0.0720 -1.1859  0.0740 -1.1951  0.0760 -1.2037  0.0780 -1.2117  0.0800 -1.2192  0.0820 -1.2262  0.0840 -1.2328  0.0860 -1.2388  0.0880 -1.2447  0.0900 -1.2501  0.0920 -1.2552  0.0940 -1.2600  0.0960 -1.2647  0.0980 -1.2689  0.1000 -1.2730  0.1020 -1.2768  0.1040 -1.2804  0.1060 -1.2839  0.1080 -1.2872  0.1100 -1.2902  0.1120 -1.2933  0.1140 -1.2959  0.1160 -1.2985  0.1180 -1.3009  0.1200 -1.3032  0.1220 -1.3055  0.1240 -1.3077  0.1260 -1.3098  0.1280 -1.3117  0.1300 -1.3135  0.1320 -1.3152  0.1340 -1.3168  0.1360 -1.3183  0.1380 -1.3197  0.1400 -1.3210  0.1420 -1.3223  0.1440 -1.3234  0.1460 -1.3246  0.1480 -1.3256  0.1500 -1.3267  0.1520 -1.3275  0.1540 -1.3285  0.1560 -1.3293  0.1580 -1.3300  0.1600 -1.3307  0.1620 -1.3313  0.1640 -1.3319  0.1660 -1.3325  0.1680 -1.3330  0.1700 -1.3334  0.1720 -1.3340  0.1740 -1.3343  0.1760 -1.3347  0.1780 -1.3351  0.1800 -1.3354  0.1820 -1.3357  0.1840 -1.3360  0.1860 -1.3362  0.1880 -1.3364  0.1900 -1.3366  0.1920 -1.3368  0.1940 -1.3368  0.1960 -1.3369  0.1980 -1.3370  0.2000 -1.3371  0.2020 -1.3371  0.2040 -1.3371  0.2060 -1.3370  0.2080 -1.3370  0.2100 -1.3370  0.2120 -1.3368  0.2140 -1.3368  0.2160 -1.3366  0.2180 -1.3365  0.2200 -1.3363  0.2220 -1.3362  0.2240 -1.3360  0.2260 -1.3358  0.2280 -1.3356  0.2300 -1.3353  0.2320 -1.3350  0.2340 -1.3348  0.2360 -1.3346  0.2380 -1.3344  0.2400 -1.3341  0.2420 -1.3337  0.2440 -1.3333  0.2460 -1.3329  0.2480 -1.3326  0.2500 -1.3323  0.2520 -1.3319  0.2540 -1.3315  0.2560 -1.3311  0.2580 -1.3307  0.2600 -1.3303  0.2620 -1.3297  0.2640 -1.3293  0.2660 -1.3289  0.2680 -1.3285  0.2700 -1.3279  0.2720 -1.3275  0.2740 -1.3269  0.2760 -1.3264  0.2780 -1.3259  0.2800 -1.3254  0.2820 -1.3249  0.2840 -1.3244  0.2860 -1.3238  0.2880 -1.3232  0.2900 -1.3226  0.2920 -1.3221  0.2940 -1.3214  0.2960 -1.3208  0.2980 -1.3202  0.3000 -1.3197  0.3020 -1.3191  0.3040 -1.3184  0.3060 -1.3179  0.3080 -1.3172  0.3100 -1.3165  0.3120 -1.3159  0.3140 -1.3153  0.3160 -1.3146  0.3180 -1.3140  0.3200 -1.3134  0.3220 -1.3127  0.3240 -1.3120  0.3260 -1.3114  0.3280 -1.3107  0.3300 -1.3099  0.3320 -1.3092  0.3340 -1.3085  0.3360 -1.3077  0.3380 -1.3070  0.3400 -1.3063  0.3420 -1.3055  0.3440 -1.3048  0.3460 -1.3042  0.3480 -1.3034  0.3500 -1.3027  0.3520 -1.3020  0.3540 -1.3013  0.3560 -1.3005  0.3580 -1.2997  0.3600 -1.2989  0.3620 -1.2982  0.3640 -1.2974  0.3660 -1.2966  0.3680 -1.2958  0.3700 -1.2950  0.3720 -1.2943  0.3740 -1.2934  0.3760 -1.2926  0.3780 -1.2919  0.3800 -1.2911  0.3820 -1.2903  0.3840 -1.2895  0.3860 -1.2887  0.3880 -1.2878  0.3900 -1.2870  0.3920 -1.2861  0.3940 -1.2853  0.3960 -1.2845  0.3980 -1.2838  0.4000 -1.2829  0.4020 -1.2821  0.4040 -1.2813  0.4060 -1.2804  0.4080 -1.2796  0.4100 -1.2789  0.4120 -1.2780  0.4140 -1.2772  0.4160 -1.2763  0.4180 -1.2754  0.4200 -1.2745  0.4220 -1.2735  0.4240 -1.2723  0.4260 -1.2714  0.4280 -1.2705  0.4300 -1.2695  0.4320 -1.2687  0.4340 -1.2679  0.4360 -1.2669  0.4380 -1.2659  0.4400 -1.2650  0.4420 -1.2641  0.4440 -1.2633  0.4460 -1.2623  0.4480 -1.2613  0.4500 -1.2603  0.4520 -1.2594  0.4540 -1.2584  0.4560 -1.2574  0.4580 -1.2564  0.4600 -1.2554  0.4620 -1.2544  0.4640 -1.2534  0.4660 -1.2524  0.4680 -1.2514  0.4700 -1.2503  0.4720 -1.2493  0.4740 -1.2482  0.4760 -1.2472  0.4780 -1.2462  0.4800 -1.2452  0.4820 -1.2441  0.4840 -1.2431  0.4860 -1.2420  0.4880 -1.2410  0.4900 -1.2399  0.4920 -1.2388  0.4940 -1.2377  0.4960 -1.2366  0.4980 -1.2356  0.5000 -1.2345  0.5020 -1.2334  0.5040 -1.2323  0.5060 -1.2312  0.5080 -1.2301  0.5100 -1.2290  0.5120 -1.2279  0.5140 -1.2268  0.5160 -1.2257  0.5180 -1.2246  0.5200 -1.2234  0.5220 -1.2223  0.5240 -1.2212  0.5260 -1.2200  0.5280 -1.2189  0.5300 -1.2177  0.5320 -1.2166  0.5340 -1.2154  0.5360 -1.2143  0.5380 -1.2131  0.5400 -1.2119  0.5420 -1.2108  0.5440 -1.2096  0.5460 -1.2084  0.5480 -1.2072  0.5500 -1.2060  0.5520 -1.2049  0.5540 -1.2036  0.5560 -1.2024  0.5580 -1.2012  0.5600 -1.2000  0.5620 -1.1988  0.5640 -1.1976  0.5660 -1.1964  0.5680 -1.1951  0.5700 -1.1939  0.5720 -1.1927  0.5740 -1.1914  0.5760 -1.1902  0.5780 -1.1889  0.5800 -1.1876  0.5820 -1.1864  0.5840 -1.1851  0.5860 -1.1839  0.5880 -1.1826  0.5900 -1.1813  0.5920 -1.1800  0.5940 -1.1787  0.5960 -1.1774  0.5980 -1.1762  0.6000 -1.1748  0.6020 -1.1735  0.6040 -1.1722  0.6060 -1.1709  0.6080 -1.1696  0.6100 -1.1682  0.6120 -1.1669  0.6140 -1.1656  0.6160 -1.1642  0.6180 -1.1629  0.6200 -1.1615  0.6220 -1.1601  0.6240 -1.1588  0.6260 -1.1574  0.6280 -1.1561  0.6300 -1.1547  0.6320 -1.1533  0.6340 -1.1519  0.6360 -1.1505  0.6380 -1.1491  0.6400 -1.1477  0.6420 -1.1463  0.6440 -1.1449  0.6460 -1.1434  0.6480 -1.1420  0.6500 -1.1405  0.6520 -1.1391  0.6540 -1.1376  0.6560 -1.1362  0.6580 -1.1347  0.6600 -1.1333  0.6620 -1.1318  0.6640 -1.1303  0.6660 -1.1288  0.6680 -1.1273  0.6700 -1.1258  0.6720 -1.1243  0.6740 -1.1228  0.6760 -1.1213  0.6780 -1.1198  0.6800 -1.1182  0.6820 -1.1167  0.6840 -1.1152  0.6860 -1.1136  0.6880 -1.1121  0.6900 -1.1105  0.6920 -1.1090  0.6940 -1.1074  0.6960 -1.1058  0.6980 -1.1043  0.7000 -1.1027  0.7020 -1.1011  0.7040 -1.0995  0.7060 -1.0979  0.7080 -1.0963  0.7100 -1.0947  0.7120 -1.0930  0.7140 -1.0914  0.7160 -1.0897  0.7180 -1.0881  0.7200 -1.0864  0.7220 -1.0848  0.7240 -1.0831  0.7260 -1.0814  0.7280 -1.0797  0.7300 -1.0780  0.7320 -1.0763  0.7340 -1.0746  0.7360 -1.0729  0.7380 -1.0712  0.7400 -1.0694  0.7420 -1.0677  0.7440 -1.0659  0.7460 -1.0642  0.7480 -1.0624  0.7500 -1.0606  0.7520 -1.0588  0.7540 -1.0570  0.7560 -1.0552  0.7580 -1.0534  0.7600 -1.0516  0.7620 -1.0498  0.7640 -1.0479  0.7660 -1.0461  0.7680 -1.0442  0.7700 -1.0423  0.7720 -1.0404  0.7740 -1.0386  0.7760 -1.0367  0.7780 -1.0347  0.7800 -1.0328  0.7820 -1.0309  0.7840 -1.0290  0.7860 -1.0270  0.7880 -1.0250  0.7900 -1.0231  0.7920 -1.0211  0.7940 -1.0191  0.7960 -1.0171  0.7980 -1.0151  0.8000 -1.0130  0.8020 -1.0110  0.8040 -1.0089  0.8060 -1.0069  0.8080 -1.0048  0.8100 -1.0027  0.8120 -1.0006  0.8140 -0.9985  0.8160 -0.9964  0.8180 -0.9943  0.8200 -0.9921  0.8220 -0.9900  0.8240 -0.9878  0.8260 -0.9856  0.8280 -0.9834  0.8300 -0.9813  0.8320 -0.9790  0.8340 -0.9768  0.8360 -0.9745  0.8380 -0.9723  0.8400 -0.9700  0.8420 -0.9677  0.8440 -0.9654  0.8460 -0.9631  0.8480 -0.9607  0.8500 -0.9584  0.8520 -0.9560  0.8540 -0.9536  0.8560 -0.9512  0.8580 -0.9488  0.8600 -0.9464  0.8620 -0.9439  0.8640 -0.9414  0.8660 -0.9389  0.8680 -0.9364  0.8700 -0.9339  0.8720 -0.9313  0.8740 -0.9287  0.8760 -0.9262  0.8780 -0.9236  0.8800 -0.9210  0.8820 -0.9183  0.8840 -0.9156  0.8860 -0.9129  0.8880 -0.9102  0.8900 -0.9075  0.8920 -0.9047  0.8940 -0.9020  0.8960 -0.8992  0.8980 -0.8963  0.9000 -0.8935  0.9020 -0.8906  0.9040 -0.8877  0.9060 -0.8847  0.9080 -0.8818  0.9100 -0.8788  0.9120 -0.8758  0.9140 -0.8727  0.9160 -0.8696  0.9180 -0.8665  0.9200 -0.8634  0.9220 -0.8602  0.9240 -0.8569  0.9260 -0.8537  0.9280 -0.8504  0.9300 -0.8470  0.9320 -0.8437  0.9340 -0.8401  0.9360 -0.8367  0.9380 -0.8332  0.9400 -0.8296  0.9420 -0.8261  0.9440 -0.8223  0.9460 -0.8186  0.9480 -0.8149  0.9500 -0.8110  0.9520 -0.8071  0.9540 -0.8032  0.9560 -0.7992  0.9580 -0.7951  0.9600 -0.7910  0.9620 -0.7868  0.9640 -0.7825  0.9660 -0.7781  0.9680 -0.7737  0.9700 -0.7692  0.9720 -0.7645  0.9740 -0.7597  0.9760 -0.7549  0.9780 -0.7499  0.9800 -0.7449  0.9820 -0.7396  0.9840 -0.7341  0.9860 -0.7282  0.9880 -0.7225  0.9900 -0.7163  0.9920 -0.7101  0.9940 -0.7034  0.9960 -0.6962  0.9980 -0.6889  1.0000 -0.6807     /
\multiput {\large$\cdot$} at 0.0300 -0.4620  0.0320 -0.5839  0.0340 -0.6739  0.0360 -0.7452  0.0380 -0.8041  0.0400 -0.8536  0.0420 -0.8960  0.0440 -0.9328  0.0460 -0.9652  0.0480 -0.9937  0.0500 -1.0191  0.0520 -1.0422  0.0540 -1.0628  0.0560 -1.0817  0.0580 -1.0987  0.0600 -1.1144  0.0620 -1.1287  0.0640 -1.1420  0.0660 -1.1544  0.0680 -1.1657  0.0700 -1.1762  0.0720 -1.1859  0.0740 -1.1951  0.0760 -1.2037  0.0780 -1.2117  0.0800 -1.2192  0.0820 -1.2262  0.0840 -1.2328  0.0860 -1.2388  0.0880 -1.2447  0.0900 -1.2501  0.0920 -1.2552  0.0940 -1.2600  0.0960 -1.2647  0.0980 -1.2689  0.1000 -1.2730  0.1020 -1.2768  0.1040 -1.2804  0.1060 -1.2839  0.1080 -1.2872  0.1100 -1.2902  0.1120 -1.2933  0.1140 -1.2959  0.1160 -1.2985  0.1180 -1.3009  0.1200 -1.3032  0.1220 -1.3055  0.1240 -1.3077  0.1260 -1.3098  0.1280 -1.3117  0.1300 -1.3135  0.1320 -1.3152  0.1340 -1.3168  0.1360 -1.3183  0.1380 -1.3197  0.1400 -1.3210  0.1420 -1.3223  0.1440 -1.3234  0.1460 -1.3246  0.1480 -1.3256  0.1500 -1.3267  0.1520 -1.3275  0.1540 -1.3285  0.1560 -1.3293  0.1580 -1.3300  0.1600 -1.3307  0.1620 -1.3313  0.1640 -1.3319  0.1660 -1.3325  0.1680 -1.3330  0.1700 -1.3334  0.1720 -1.3340  0.1740 -1.3343  0.1760 -1.3347  0.1780 -1.3351  0.1800 -1.3354  0.1820 -1.3357  0.1840 -1.3360  0.1860 -1.3362  0.1880 -1.3364  0.1900 -1.3366  0.1920 -1.3368  0.1940 -1.3368  0.1960 -1.3369  0.1980 -1.3370  0.2000 -1.3371  0.2020 -1.3371  0.2040 -1.3371  0.2060 -1.3370  0.2080 -1.3370  0.2100 -1.3370  0.2120 -1.3368  0.2140 -1.3368  0.2160 -1.3366  0.2180 -1.3365  0.2200 -1.3363  0.2220 -1.3362  0.2240 -1.3360  0.2260 -1.3358  0.2280 -1.3356  0.2300 -1.3353  0.2320 -1.3350  0.2340 -1.3348  0.2360 -1.3346  0.2380 -1.3344  0.2400 -1.3341  0.2420 -1.3337  0.2440 -1.3333  0.2460 -1.3329  0.2480 -1.3326  0.2500 -1.3323  0.2520 -1.3319  0.2540 -1.3315  0.2560 -1.3311  0.2580 -1.3307  0.2600 -1.3303  0.2620 -1.3297  0.2640 -1.3293  0.2660 -1.3289  0.2680 -1.3285  0.2700 -1.3279  0.2720 -1.3275  0.2740 -1.3269  0.2760 -1.3264  0.2780 -1.3259  0.2800 -1.3254  0.2820 -1.3249  0.2840 -1.3244  0.2860 -1.3238  0.2880 -1.3232  0.2900 -1.3226  0.2920 -1.3221  0.2940 -1.3214  0.2960 -1.3208  0.2980 -1.3202  0.3000 -1.3197  0.3020 -1.3191  0.3040 -1.3184  0.3060 -1.3179  0.3080 -1.3172  0.3100 -1.3165  0.3120 -1.3159  0.3140 -1.3153  0.3160 -1.3146  0.3180 -1.3140  0.3200 -1.3134  0.3220 -1.3127  0.3240 -1.3120  0.3260 -1.3114  0.3280 -1.3107  0.3300 -1.3099  0.3320 -1.3092  0.3340 -1.3085  0.3360 -1.3077  0.3380 -1.3070  0.3400 -1.3063  0.3420 -1.3055  0.3440 -1.3048  0.3460 -1.3042  0.3480 -1.3034  0.3500 -1.3027  0.3520 -1.3020  0.3540 -1.3013  0.3560 -1.3005  0.3580 -1.2997  0.3600 -1.2989  0.3620 -1.2982  0.3640 -1.2974  0.3660 -1.2966  0.3680 -1.2958  0.3700 -1.2950  0.3720 -1.2943  0.3740 -1.2934  0.3760 -1.2926  0.3780 -1.2919  0.3800 -1.2911  0.3820 -1.2903  0.3840 -1.2895  0.3860 -1.2887  0.3880 -1.2878  0.3900 -1.2870  0.3920 -1.2861  0.3940 -1.2853  0.3960 -1.2845  0.3980 -1.2838  0.4000 -1.2829  0.4020 -1.2821  0.4040 -1.2813  0.4060 -1.2804  0.4080 -1.2796  0.4100 -1.2789  0.4120 -1.2780  0.4140 -1.2772  0.4160 -1.2763  0.4180 -1.2754  0.4200 -1.2745  0.4220 -1.2735  0.4240 -1.2723  0.4260 -1.2714  0.4280 -1.2705  0.4300 -1.2695  0.4320 -1.2687  0.4340 -1.2679  0.4360 -1.2669  0.4380 -1.2659  0.4400 -1.2650  0.4420 -1.2641  0.4440 -1.2633  0.4460 -1.2623  0.4480 -1.2613  0.4500 -1.2603  0.4520 -1.2594  0.4540 -1.2584  0.4560 -1.2574  0.4580 -1.2564  0.4600 -1.2554  0.4620 -1.2544  0.4640 -1.2534  0.4660 -1.2524  0.4680 -1.2514  0.4700 -1.2503  0.4720 -1.2493  0.4740 -1.2482  0.4760 -1.2472  0.4780 -1.2462  0.4800 -1.2452  0.4820 -1.2441  0.4840 -1.2431  0.4860 -1.2420  0.4880 -1.2410  0.4900 -1.2399  0.4920 -1.2388  0.4940 -1.2377  0.4960 -1.2366  0.4980 -1.2356  0.5000 -1.2345  0.5020 -1.2334  0.5040 -1.2323  0.5060 -1.2312  0.5080 -1.2301  0.5100 -1.2290  0.5120 -1.2279  0.5140 -1.2268  0.5160 -1.2257  0.5180 -1.2246  0.5200 -1.2234  0.5220 -1.2223  0.5240 -1.2212  0.5260 -1.2200  0.5280 -1.2189  0.5300 -1.2177  0.5320 -1.2166  0.5340 -1.2154  0.5360 -1.2143  0.5380 -1.2131  0.5400 -1.2119  0.5420 -1.2108  0.5440 -1.2096  0.5460 -1.2084  0.5480 -1.2072  0.5500 -1.2060  0.5520 -1.2049  0.5540 -1.2036  0.5560 -1.2024  0.5580 -1.2012  0.5600 -1.2000  0.5620 -1.1988  0.5640 -1.1976  0.5660 -1.1964  0.5680 -1.1951  0.5700 -1.1939  0.5720 -1.1927  0.5740 -1.1914  0.5760 -1.1902  0.5780 -1.1889  0.5800 -1.1876  0.5820 -1.1864  0.5840 -1.1851  0.5860 -1.1839  0.5880 -1.1826  0.5900 -1.1813  0.5920 -1.1800  0.5940 -1.1787  0.5960 -1.1774  0.5980 -1.1762  0.6000 -1.1748  0.6020 -1.1735  0.6040 -1.1722  0.6060 -1.1709  0.6080 -1.1696  0.6100 -1.1682  0.6120 -1.1669  0.6140 -1.1656  0.6160 -1.1642  0.6180 -1.1629  0.6200 -1.1615  0.6220 -1.1601  0.6240 -1.1588  0.6260 -1.1574  0.6280 -1.1561  0.6300 -1.1547  0.6320 -1.1533  0.6340 -1.1519  0.6360 -1.1505  0.6380 -1.1491  0.6400 -1.1477  0.6420 -1.1463  0.6440 -1.1449  0.6460 -1.1434  0.6480 -1.1420  0.6500 -1.1405  0.6520 -1.1391  0.6540 -1.1376  0.6560 -1.1362  0.6580 -1.1347  0.6600 -1.1333  0.6620 -1.1318  0.6640 -1.1303  0.6660 -1.1288  0.6680 -1.1273  0.6700 -1.1258  0.6720 -1.1243  0.6740 -1.1228  0.6760 -1.1213  0.6780 -1.1198  0.6800 -1.1182  0.6820 -1.1167  0.6840 -1.1152  0.6860 -1.1136  0.6880 -1.1121  0.6900 -1.1105  0.6920 -1.1090  0.6940 -1.1074  0.6960 -1.1058  0.6980 -1.1043  0.7000 -1.1027  0.7020 -1.1011  0.7040 -1.0995  0.7060 -1.0979  0.7080 -1.0963  0.7100 -1.0947  0.7120 -1.0930  0.7140 -1.0914  0.7160 -1.0897  0.7180 -1.0881  0.7200 -1.0864  0.7220 -1.0848  0.7240 -1.0831  0.7260 -1.0814  0.7280 -1.0797  0.7300 -1.0780  0.7320 -1.0763  0.7340 -1.0746  0.7360 -1.0729  0.7380 -1.0712  0.7400 -1.0694  0.7420 -1.0677  0.7440 -1.0659  0.7460 -1.0642  0.7480 -1.0624  0.7500 -1.0606  0.7520 -1.0588  0.7540 -1.0570  0.7560 -1.0552  0.7580 -1.0534  0.7600 -1.0516  0.7620 -1.0498  0.7640 -1.0479  0.7660 -1.0461  0.7680 -1.0442  0.7700 -1.0423  0.7720 -1.0404  0.7740 -1.0386  0.7760 -1.0367  0.7780 -1.0347  0.7800 -1.0328  0.7820 -1.0309  0.7840 -1.0290  0.7860 -1.0270  0.7880 -1.0250  0.7900 -1.0231  0.7920 -1.0211  0.7940 -1.0191  0.7960 -1.0171  0.7980 -1.0151  0.8000 -1.0130  0.8020 -1.0110  0.8040 -1.0089  0.8060 -1.0069  0.8080 -1.0048  0.8100 -1.0027  0.8120 -1.0006  0.8140 -0.9985  0.8160 -0.9964  0.8180 -0.9943  0.8200 -0.9921  0.8220 -0.9900  0.8240 -0.9878  0.8260 -0.9856  0.8280 -0.9834  0.8300 -0.9813  0.8320 -0.9790  0.8340 -0.9768  0.8360 -0.9745  0.8380 -0.9723  0.8400 -0.9700  0.8420 -0.9677  0.8440 -0.9654  0.8460 -0.9631  0.8480 -0.9607  0.8500 -0.9584  0.8520 -0.9560  0.8540 -0.9536  0.8560 -0.9512  0.8580 -0.9488  0.8600 -0.9464  0.8620 -0.9439  0.8640 -0.9414  0.8660 -0.9389  0.8680 -0.9364  0.8700 -0.9339  0.8720 -0.9313  0.8740 -0.9287  0.8760 -0.9262  0.8780 -0.9236  0.8800 -0.9210  0.8820 -0.9183  0.8840 -0.9156  0.8860 -0.9129  0.8880 -0.9102  0.8900 -0.9075  0.8920 -0.9047  0.8940 -0.9020  0.8960 -0.8992  0.8980 -0.8963  0.9000 -0.8935  0.9020 -0.8906  0.9040 -0.8877  0.9060 -0.8847  0.9080 -0.8818  0.9100 -0.8788  0.9120 -0.8758  0.9140 -0.8727  0.9160 -0.8696  0.9180 -0.8665  0.9200 -0.8634  0.9220 -0.8602  0.9240 -0.8569  0.9260 -0.8537  0.9280 -0.8504  0.9300 -0.8470  0.9320 -0.8437  0.9340 -0.8401  0.9360 -0.8367  0.9380 -0.8332  0.9400 -0.8296  0.9420 -0.8261  0.9440 -0.8223  0.9460 -0.8186  0.9480 -0.8149  0.9500 -0.8110  0.9520 -0.8071  0.9540 -0.8032  0.9560 -0.7992  0.9580 -0.7951  0.9600 -0.7910  0.9620 -0.7868  0.9640 -0.7825  0.9660 -0.7781  0.9680 -0.7737  0.9700 -0.7692  0.9720 -0.7645  0.9740 -0.7597  0.9760 -0.7549  0.9780 -0.7499  0.9800 -0.7449  0.9820 -0.7396  0.9840 -0.7341  0.9860 -0.7282  0.9880 -0.7225  0.9900 -0.7163  0.9920 -0.7101  0.9940 -0.7034  0.9960 -0.6962  0.9980 -0.6889  1.0000 -0.6807     /

\normalcolor
\color{black}

\endpicture